\documentclass[11pt]{article}

\usepackage[top=1in, bottom=1in, left=1in, right=1in]{geometry}

\usepackage[]{graphicx,amsmath,amssymb,amsthm,color,url}
\usepackage{psfrag}
\usepackage{subfigure}
\usepackage{algorithm}
\usepackage{algorithmic}
\usepackage{bm}

\usepackage{cite}

\begin{document}

\title{Mathematics for cryo-electron microscopy \\ \vspace{0.3cm} \large {\em To appear in the Proceedings of the International Congress of Mathematicians 2018}}

\author{Amit Singer\footnote{Department of Mathematics, and Program in Applied and Computational Mathematics, Princeton University. E-mail: \texttt{amits@math.princeton.edu}. Partially supported by Award Number R01GM090200 from the NIGMS, FA9550-17-1-0291 from AFOSR, Simons Foundation Math+X Investigator Award, and the Moore Foundation Data-Driven Discovery Investigator Award.}}
\date{}
\maketitle

\begin{abstract}
Single-particle cryo-electron microscopy (cryo-EM) has recently joined X-ray crystallography and NMR spectroscopy as a high-resolution structural method for biological macromolecules. Cryo-EM was selected by Nature Methods as Method of the Year 2015, large scale investments in cryo-EM facilities are being made all over the world, and the Nobel Prize in Chemistry 2017 was awarded to Jacques Dubochet, Joachim Frank and Richard Henderson ``for developing cryo-electron microscopy for the high-resolution structure determination of biomolecules in solution''. This paper focuses on the mathematical principles underlying existing algorithms for structure determination using single particle cryo-EM.

\end{abstract}

\section{Introduction}

The field of structural biology is currently undergoing a transformative change \cite{kuhlbrandt,smith2014beyond}. Structures of many biomolecular targets previously insurmountable by X-ray crystallography are now being obtained using single particle cryo-EM to resolutions beyond 4\AA\ on a regular basis \cite{cheng-trpv1,amunts2014structure,bartesaghi20152}. This leap in cryo-EM technology, as recognized by the 2017 Nobel Prize in Chemistry, is mainly due to hardware advancements including the invention of the direct electron detector and the methodological development of algorithms for data processing. Cryo-EM is a very general and powerful technique because it does not require the formation of crystalline arrays of macromolecules. In addition, unlike X-ray crystallography and nuclear magnetic resonance (NMR) that measure ensembles of particles, single particle cryo-EM produces images of individual particles. Cryo-EM therefore has the potential to analyze conformational changes and energy landscapes associated with structures of complexes in different functional states.

The main purpose of this brief review paper is to expose mathematicians to the exciting field of cryo-EM. As there exist many excellent review articles and textbooks on single particle cryo-EM \cite{frank,van2000single,nogales2016development,glaeser2016good,subramaniam2016cryoem,sorzano2017challenges,sigworth2016principles}, we choose to solely focus here on the mathematical foundations of this technique. Topics of great importance to practitioners, such as the physics and optics of the electron microscope, sample preparation, and data acquisition are not treated here.


In cryo-EM, biological macromolecules are imaged in an electron microscope.  The molecules are rapidly frozen in a thin layer of vitreous ice, trapping them in a nearly-physiological state. The molecules are randomly oriented and positioned within the ice layer. The electron microscope produces a two-dimensional tomographic projection image (called a micrograph) of the molecules embedded in the ice layer. More specifically, what is being measured by the detector is the integral in the direction of the beaming electrons of the electrostatic potential of the individual molecules.

Cryo-EM images, however, have very low contrast, due to the absence of heavy-metal stains or other contrast enhancements, and have very high noise due to the small electron doses that can be applied to the specimen without causing too much radiation damage. The first step in the computational pipeline is to select ``particles'' from the micrographs, that is, to crop from each micrograph several small size images each containing a single projection image, ideally centered. The molecule orientations associated with the particle images are unknown. In addition, particle images are not perfectly centered, but this would be of lesser concern to us for now.

The imaging modality is akin to the parallel beam model in Computerized Tomography (CT) of medical images, where a three-dimensional density map of an organ needs to be estimated from tomographic images. There are two aspects that make single particle reconstruction (SPR) from cryo-EM more challenging compared to classical CT. First, in medical imaging the patient avoids movement, hence viewing directions of individual projections are known to the scanning device, whereas in cryo-EM the viewing directions are unknown. Electron Tomography (ET) employs tilting and is often used for cellular imaging, providing reconstructions of lower resolution due to increased radiation damage for the entire tilt series. While it is possible to tilt the specimen and register relative viewing directions among images within a tilt series, radiation damage destroys high frequency content and it is much more difficult to obtain high resolution reconstructions using ET. In SPR, each particle image corresponds to a different molecule, ideally of the same structure, but at different and unknown orientation. Second, the signal-to-noise ratio (SNR) typical of cryo-EM images is smaller than one (more noise than signal). Thus, to obtain a reliable three-dimensional density map of a molecule, the information from many images of identical molecules must be combined.

\section{Image formation model and inverse problems}

\begin{figure}
\centering
\includegraphics[width=0.4\textwidth]{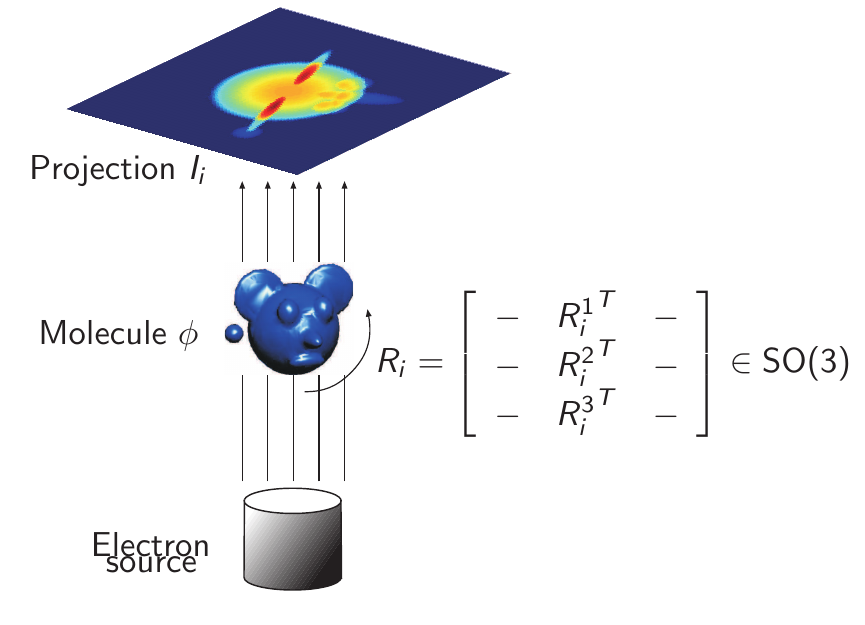}
\caption{Schematic drawing of the imaging process: every projection image corresponds to
some unknown rotation of the unknown molecule. The effect of the point spread function is not shown here.}
\label{fig:model}
\end{figure}

The mathematical image formation model is as follows (Figure \ref{fig:model}).
Let $\phi : \mathbb{R}^3 \to \mathbb{R}$ be the electrostatic potential of the molecule. Suppose that following the step of particle picking, the dataset contains $n$ particle images, denoted $I_1,\ldots, I_n$. The image $I_i$ is formed by first rotating $\phi$ by a rotation $R_i$ in $SO(3)$, then projecting the rotated molecule in the $z$-direction, convolving it with a point spread function $H_i$, sampling on a Cartesian grid of pixels of size $L\times L$, and contaminating with noise:
\begin{equation}
I_i(x,y) = H_i \star \int_{-\infty}^\infty \phi(R_i^T r) \,dz + \text{``noise"},\quad r=(x,y,z)^T.
\label{eq:forward}
\end{equation}
The rotations $R_1,\ldots, R_n \in SO(3)$ are unknown. The Fourier transform of the point spread function is called the contrast transfer function (CTF), and it is typically known, or can be estimated from the data, at least approximately, although it may vary from one image to another. Equivalently, we may rewrite the forward model (\ref{eq:forward}) as
\begin{equation}
I_i = H_i \star P R \circ \phi + \text{``noise"},
\label{eq:forward2} 
\end{equation}
where $R\circ \phi(r) = \phi(R^T r)$ and $P$ is the tomographic projection operator in the $z$-direction, $P f(x,y) = \int_{\mathbb{R}} f(x,y,z)\,dz$. We write ``noise" in Eqs.(\ref{eq:forward}) and (\ref{eq:forward2}) as a full discussion of the noise statistics and its possible dependence on the structure itself (i.e., structural noise) are beyond the scope of this paper. 

The basic cryo-EM inverse problem, called the cryo-EM reconstruction problem, is to estimate $\phi$ given $I_1,\ldots,I_n$ and $H_1,\ldots,H_n$, without knowing $R_1,\ldots,R_n$. Notice that cryo-EM reconstruction is a non-linear inverse problem, because the rotations are unknown; if the rotations were known, then it would become a linear inverse problem, for which there exist many classical solvers. Because images are finitely sampled, $\phi$ cannot be estimated beyond the resolution of the input images.

An even more challenging inverse problem is the so-called heterogeneity cryo-EM problem. Here each image may originate from a different molecular structure corresponding to possible structural variations. That is, to each image $I_i$ there may correspond a different molecular structure $\phi_i$. The goal is then to estimate $\phi_1,\ldots,\phi_n$ from $I_1,\ldots,I_n$, again, without knowing the rotations $R_1,\ldots,R_n$. Clearly, as stated, this is an ill-posed inverse problem, since we are required to estimate more output parameters (three-dimensional structures) than input data (two-dimensional images). In order to have any hope of making progress with this problem, we would need to make some restrictive assumptions about the potential functions $\phi_1,\ldots,\phi_n$. For example, the assumption of discrete variability implies that there is only a finite number of distinct conformations from which the potential functions are sampled from. Then, the goal is to estimate the number of conformations, the conformations themselves, and their distribution. Another popular assumption is that of continuous variability with a small number of flexible motions, so that $\phi_1,\ldots,\phi_n$ are sampled from a low-dimensional manifold of conformations. Either way, the problem is potentially well-posed only by assuming an underlying low-dimensional structure on the distribution of possible conformations.

In order to make this exposition less technical, we are going to make an unrealistic assumption of ideally localized point spread functions, or equivalently, constant contrast transfer functions, so that $H_1,\ldots,H_n$ are eliminated from all further consideration here. All methods and analyses considered below can be generalized to include the effect of non-ideal CTFs, unless specifically mentioned otherwise.

\section{Solving the basic cryo-EM inverse problem for clean images}

\begin{figure}
\centering
\includegraphics[width=1.6in]{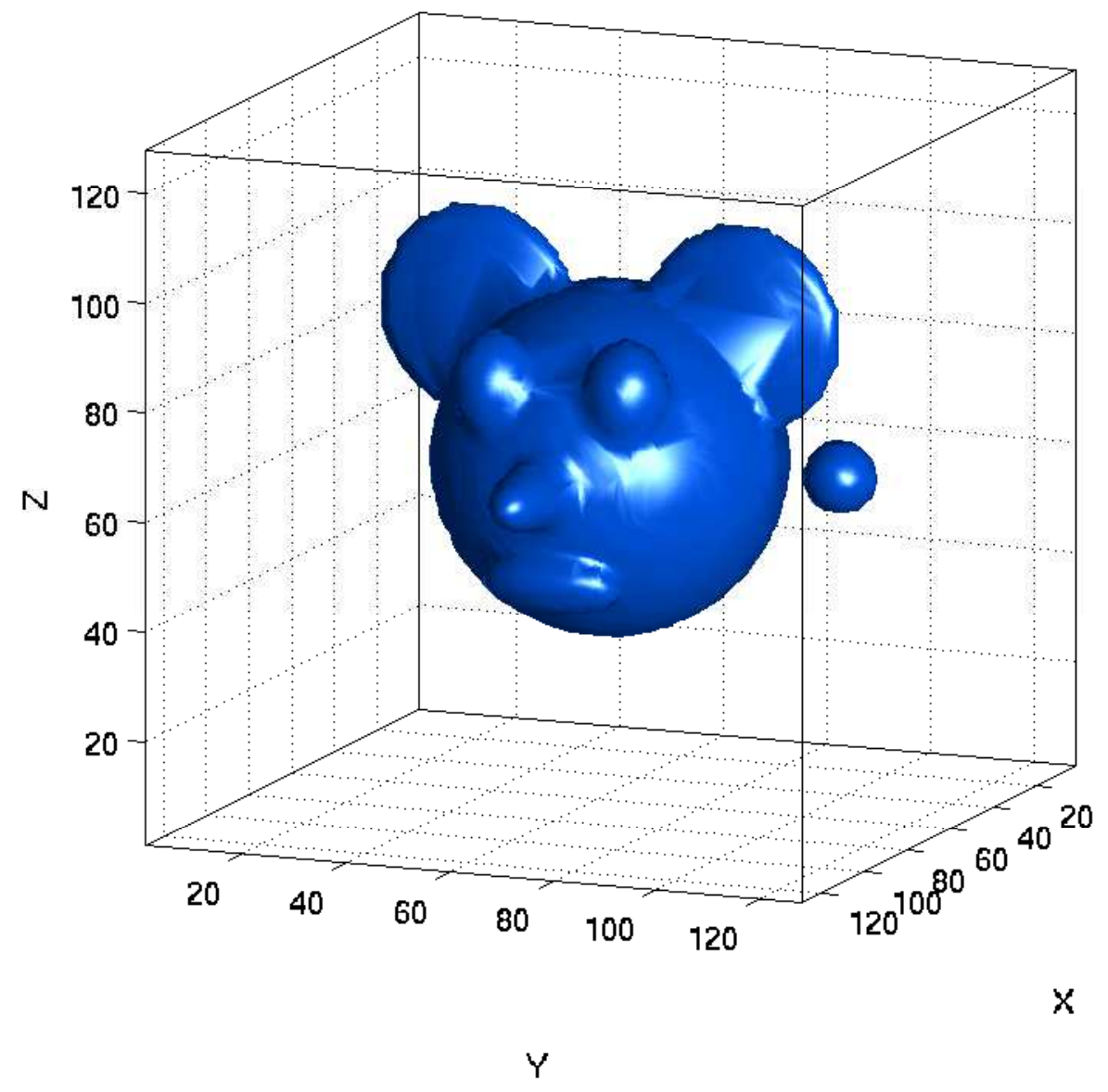}
\\
\includegraphics[width=0.6in]{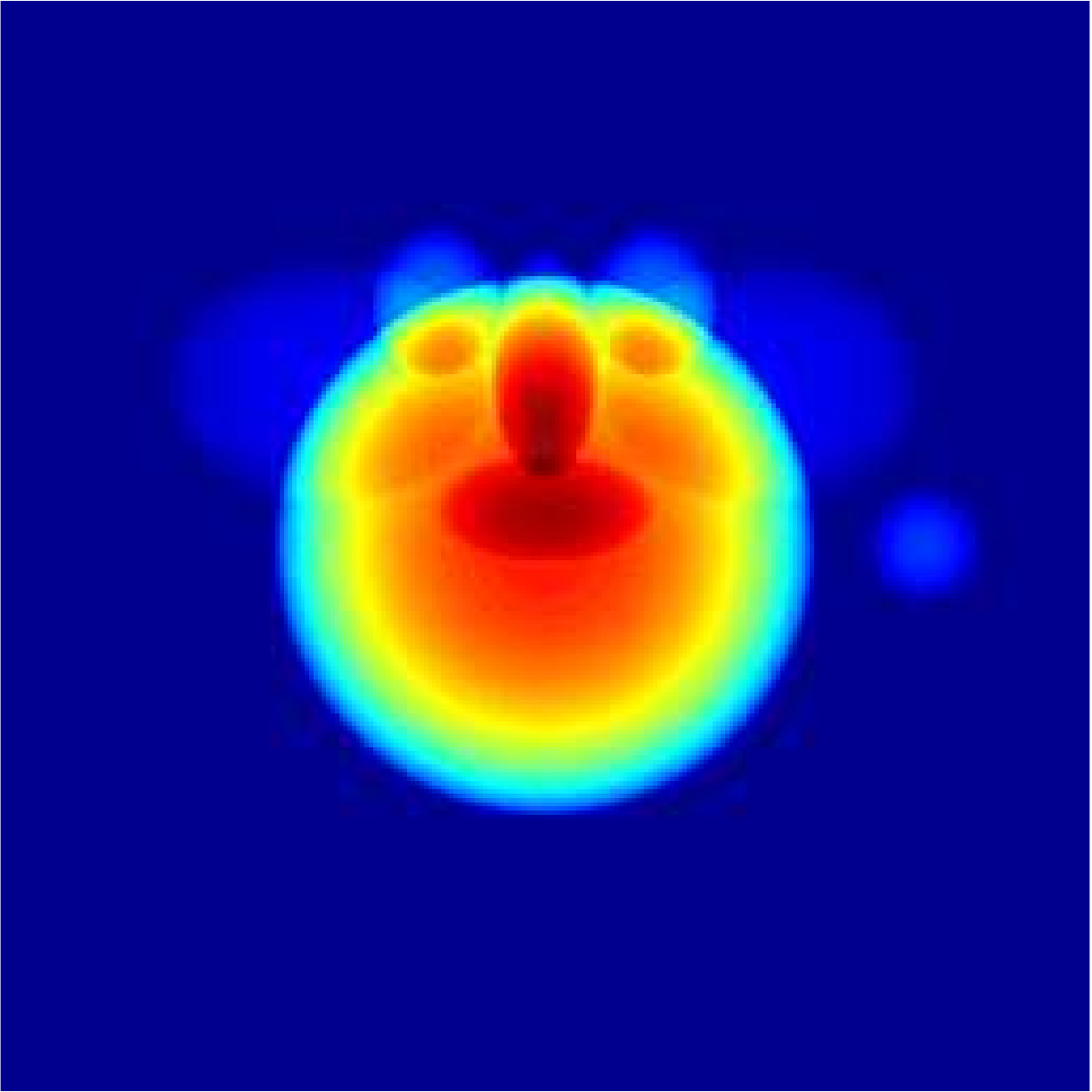}
\includegraphics[width=0.6in]{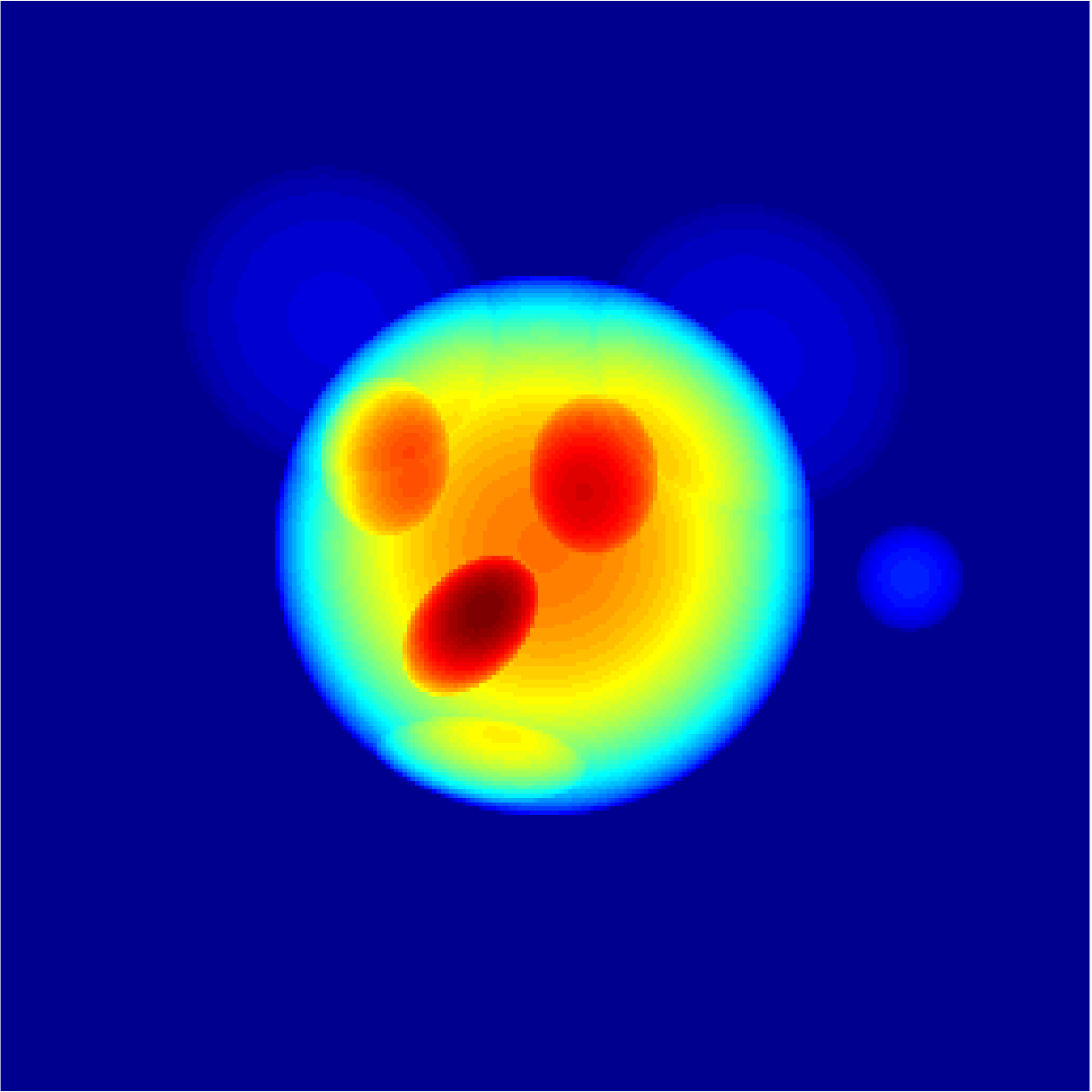}
\includegraphics[width=0.6in]{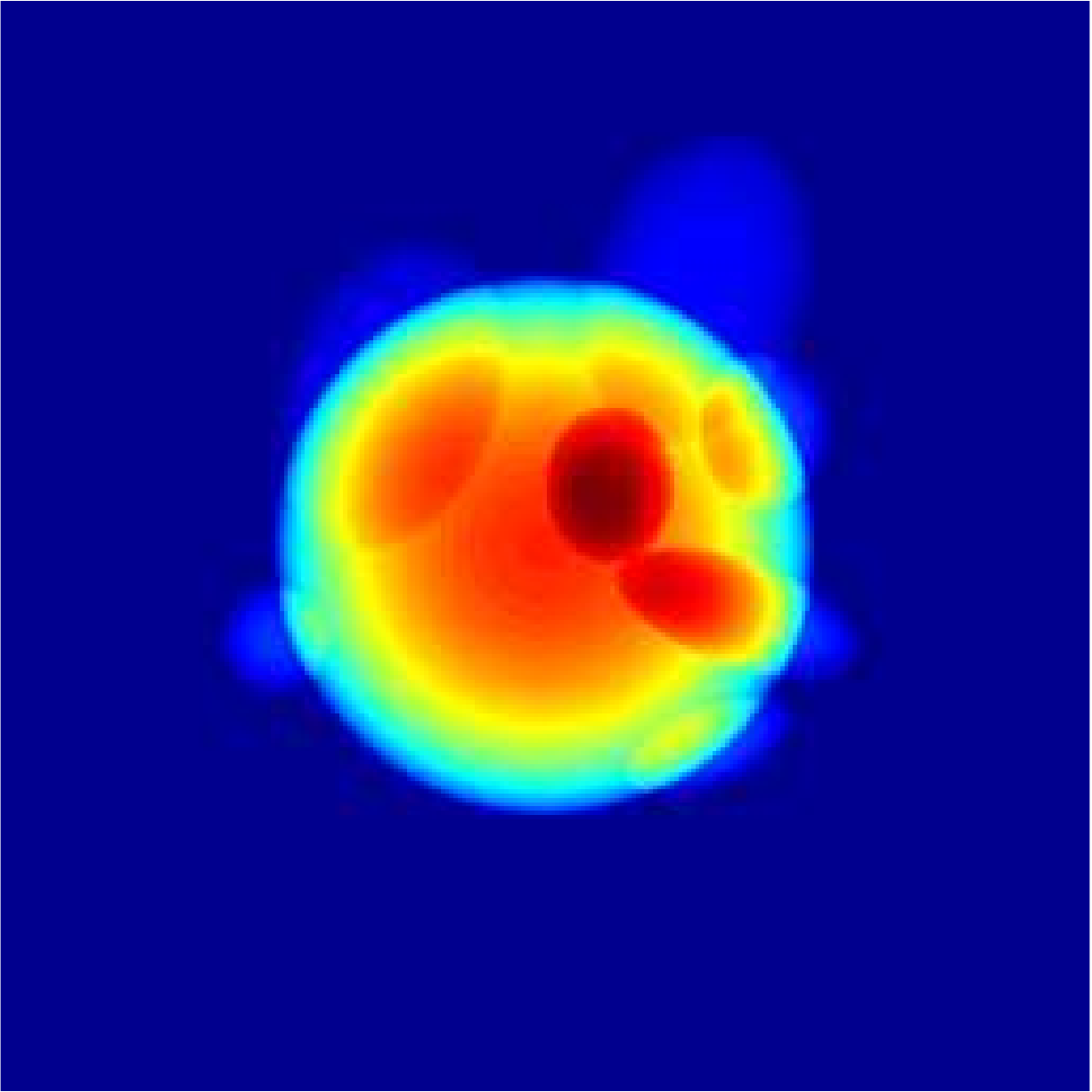}
\includegraphics[width=0.6in]{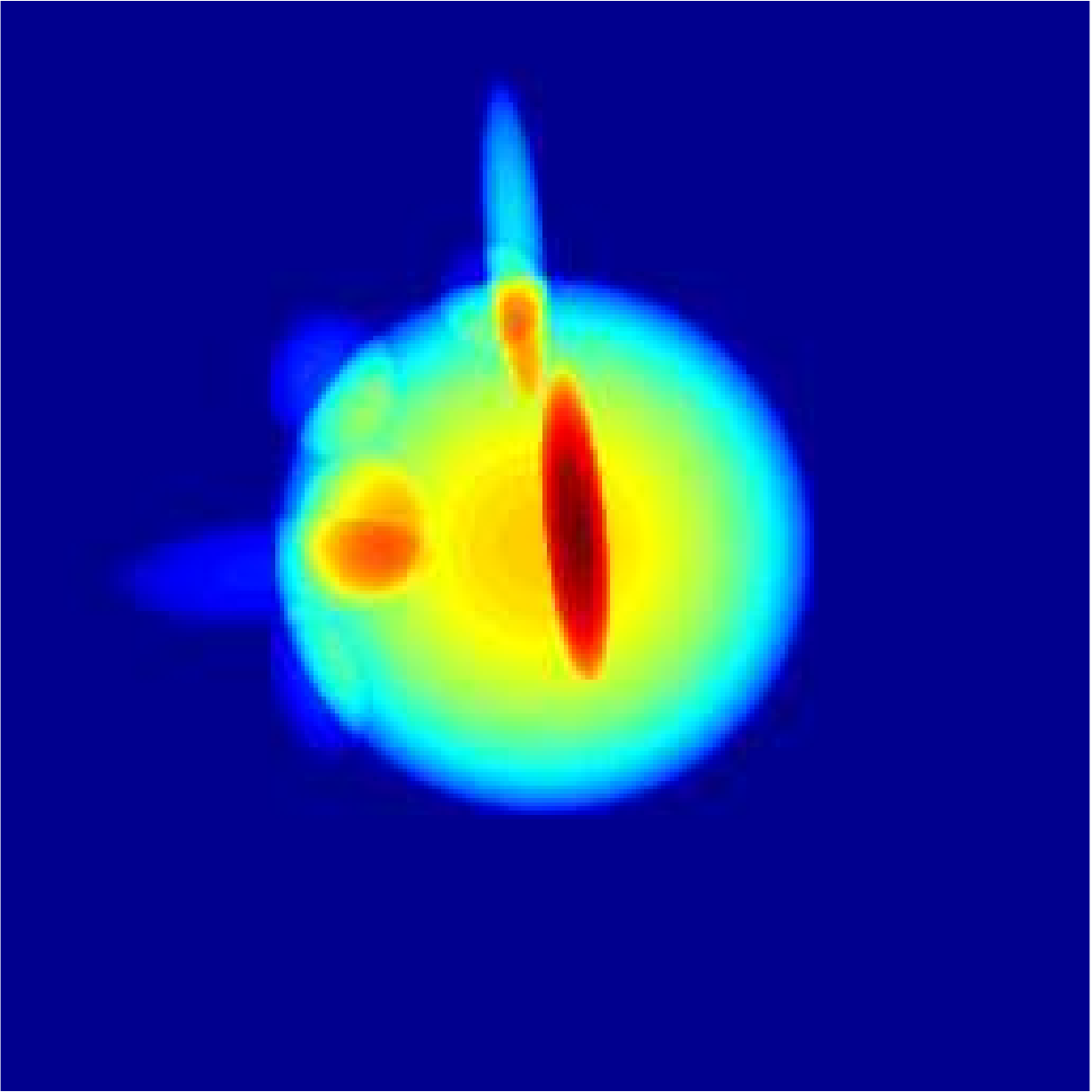}
\includegraphics[width=0.6in]{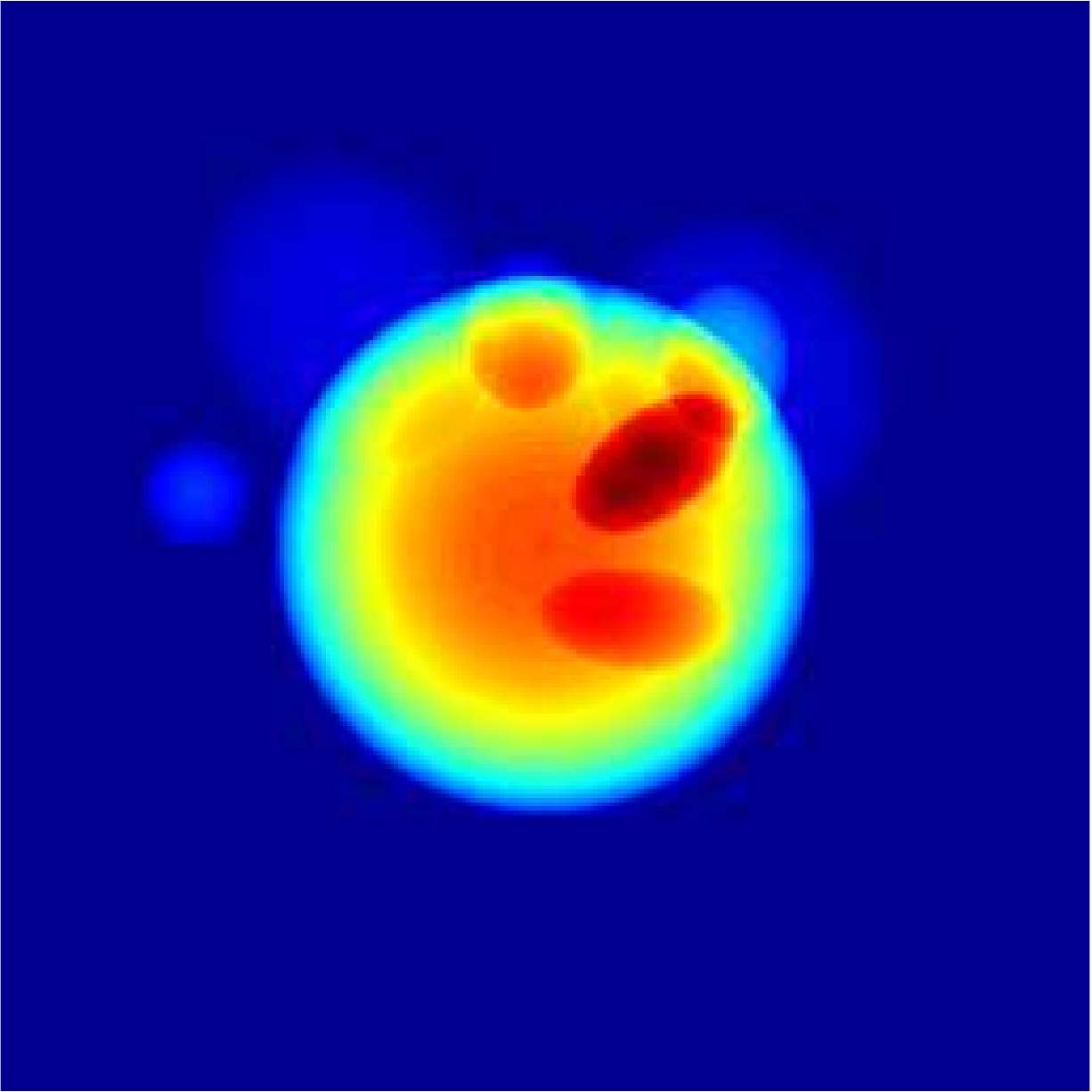}
\includegraphics[width=0.6in]{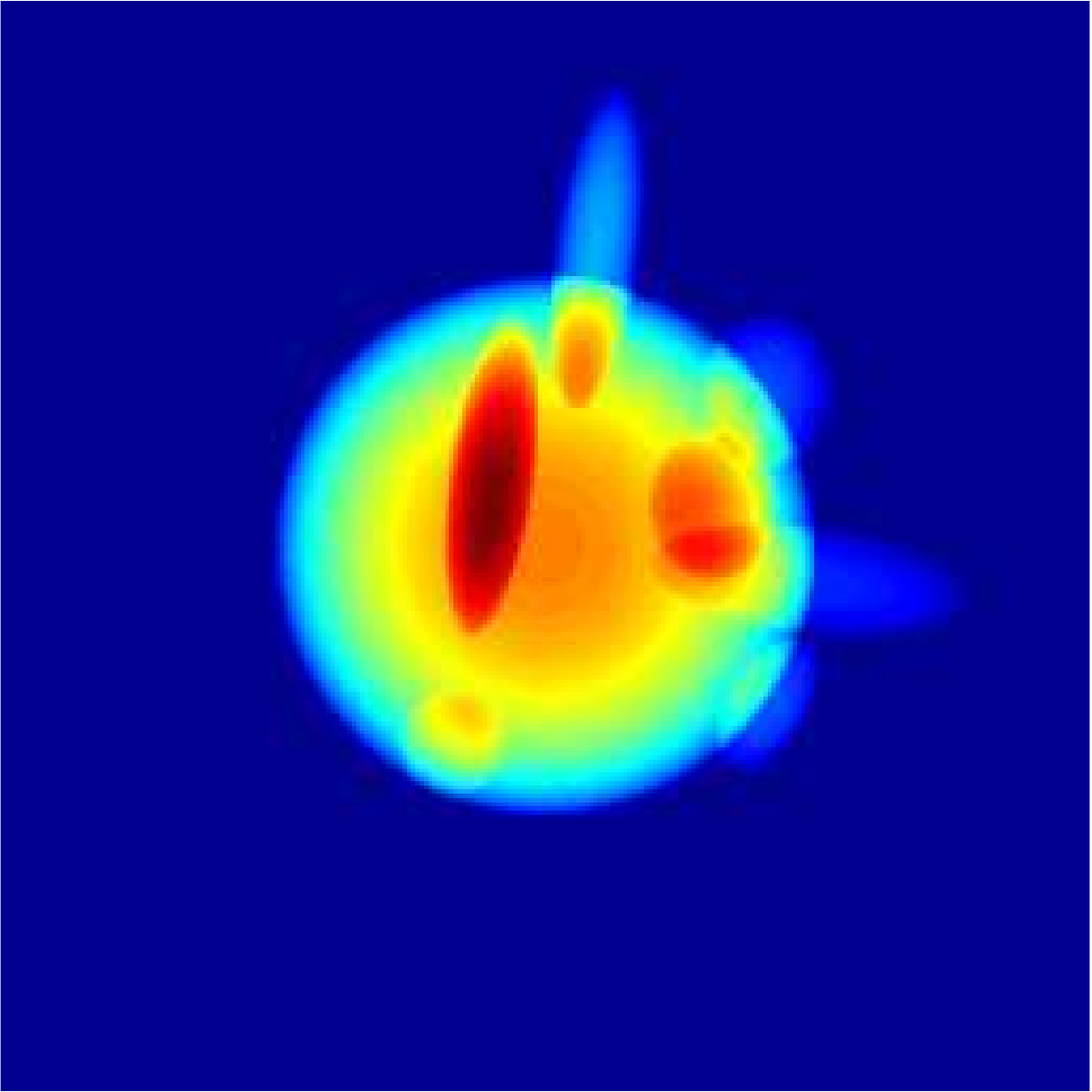}
\\
\includegraphics[width=0.6in]{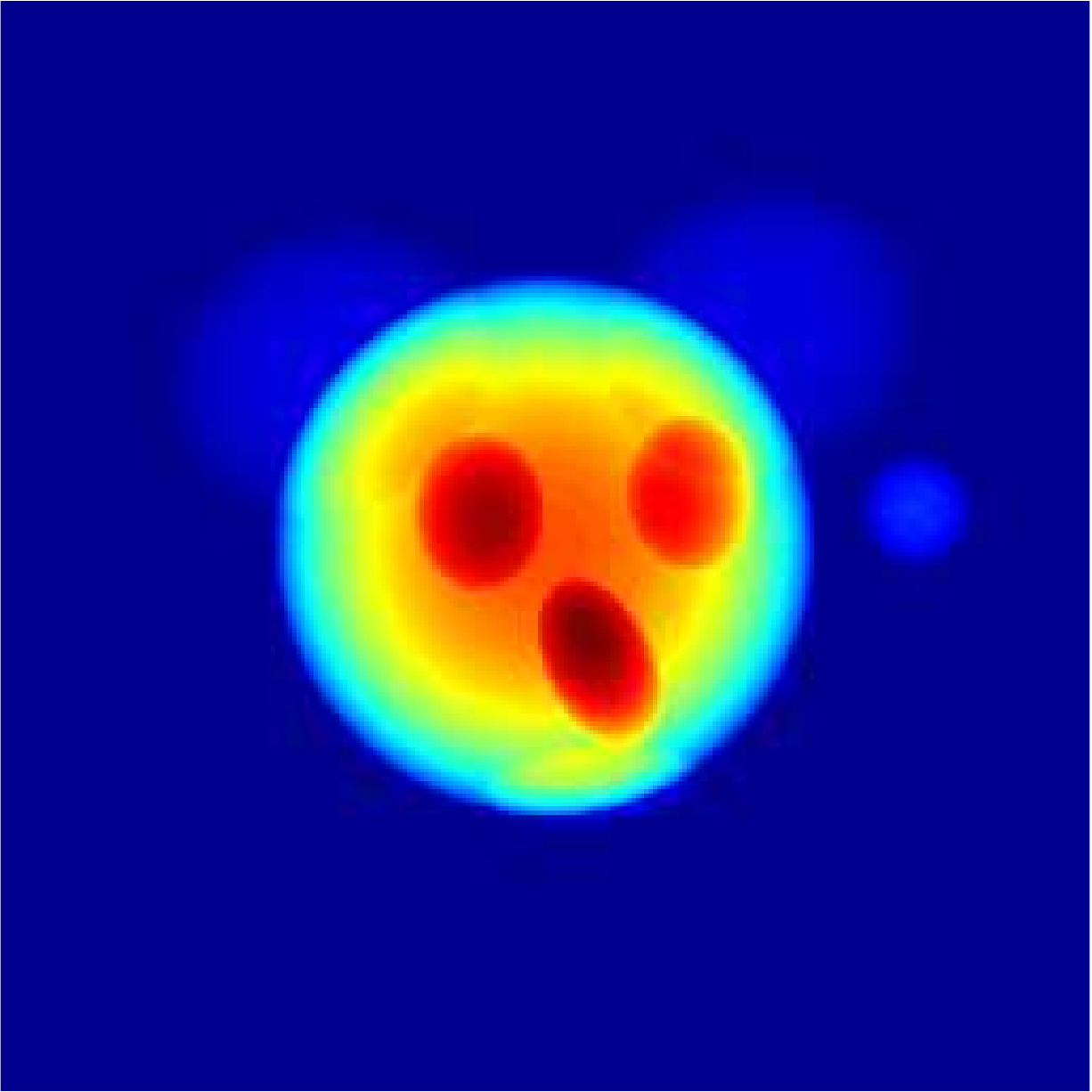}
\includegraphics[width=0.6in]{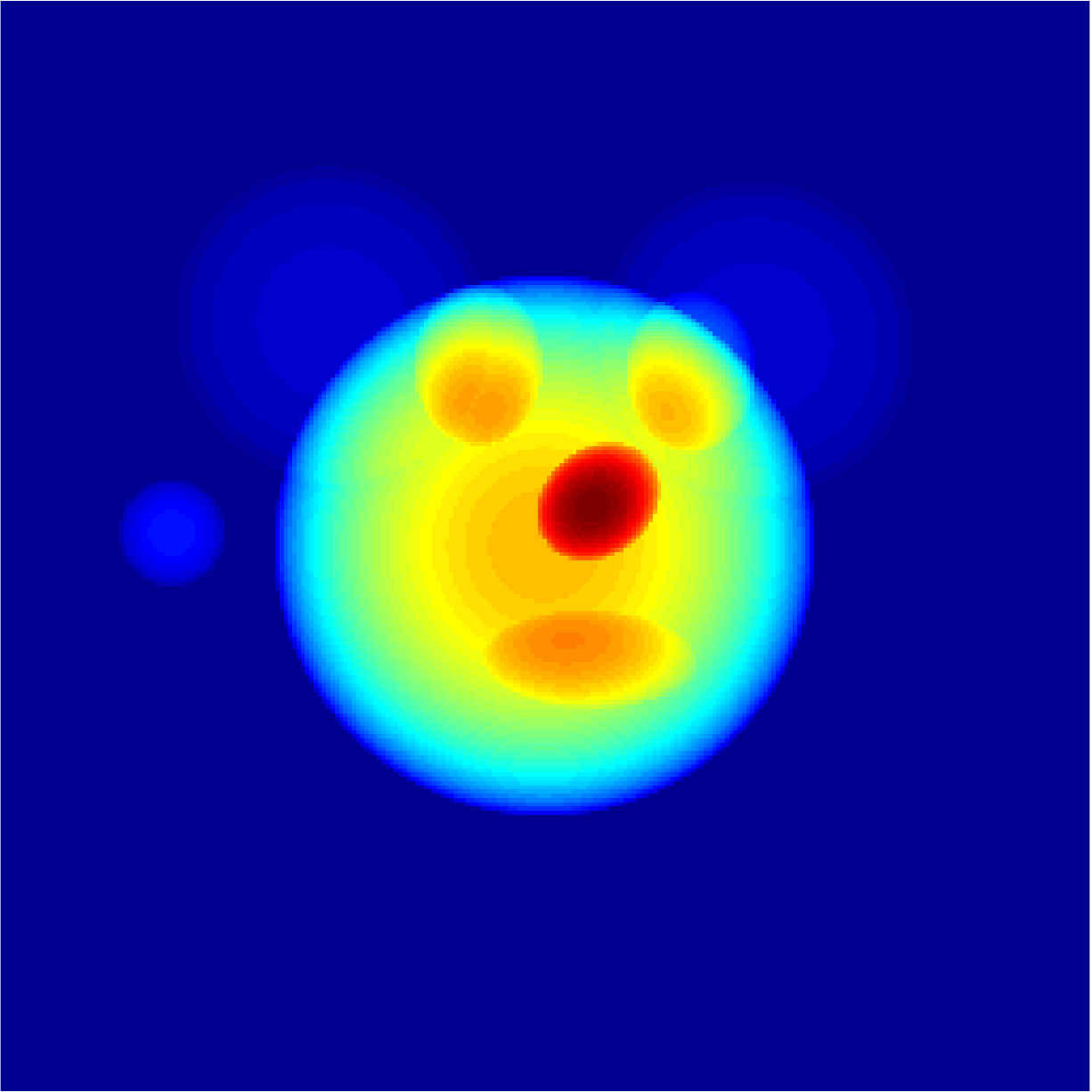}
\includegraphics[width=0.6in]{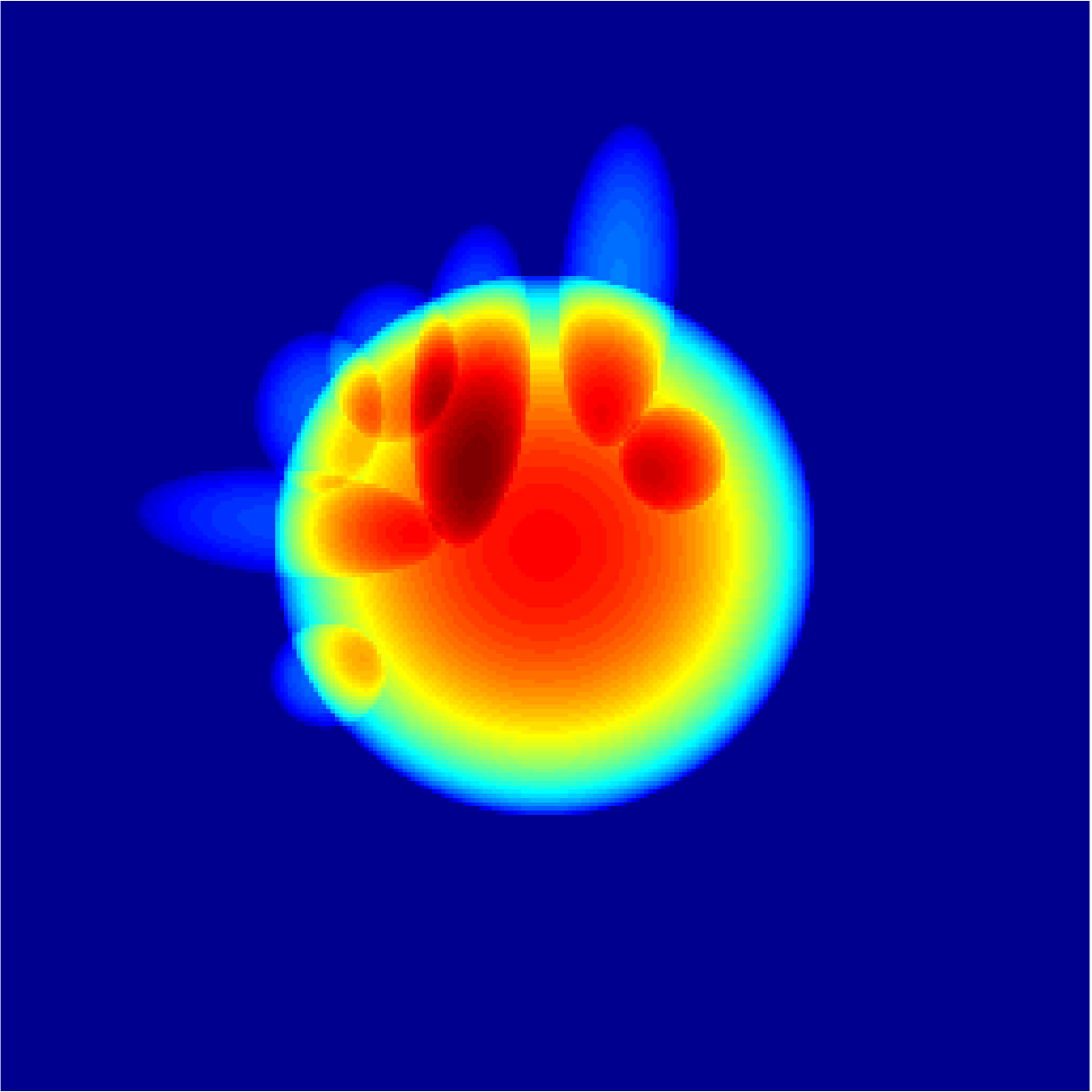}
\includegraphics[width=0.6in]{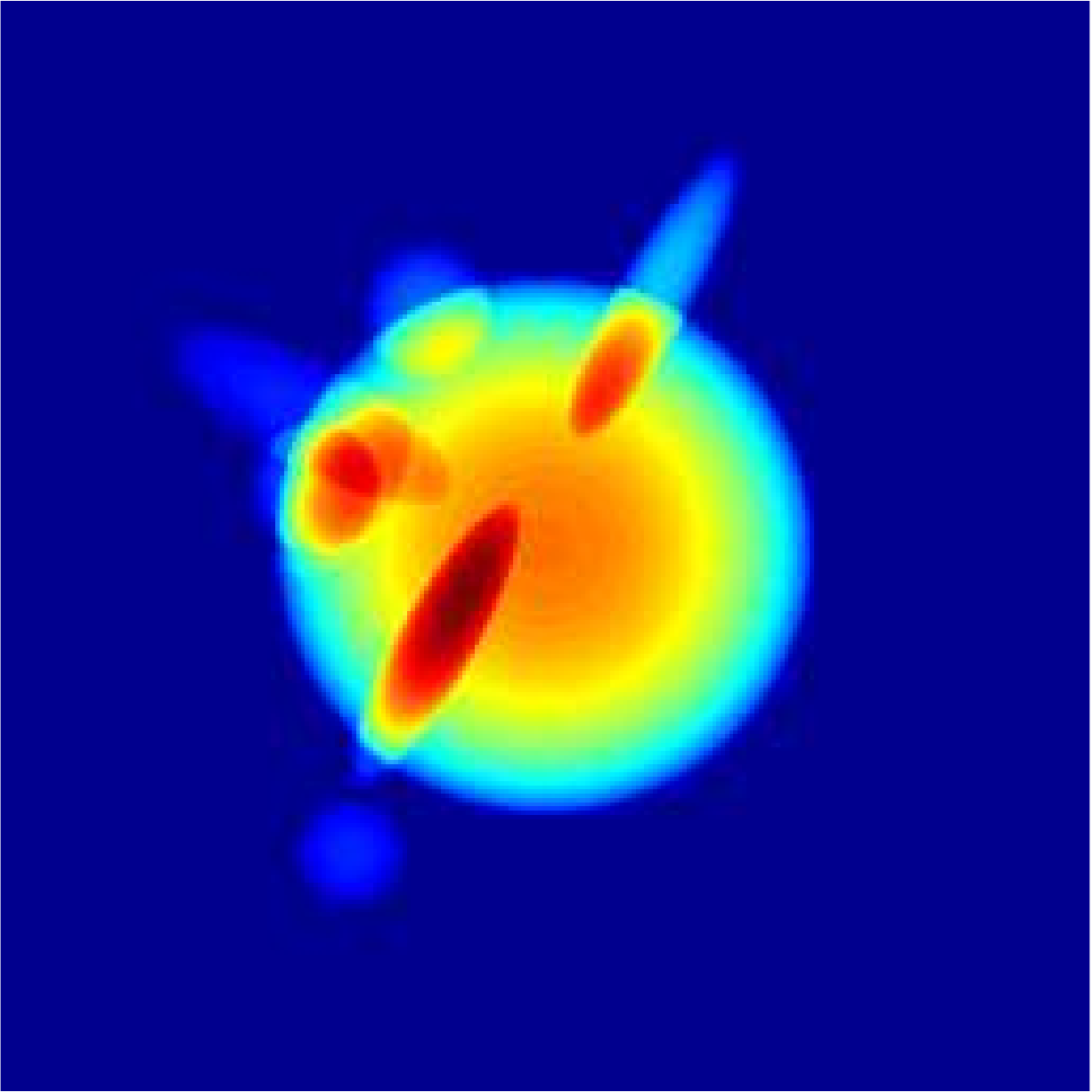}
\includegraphics[width=0.6in]{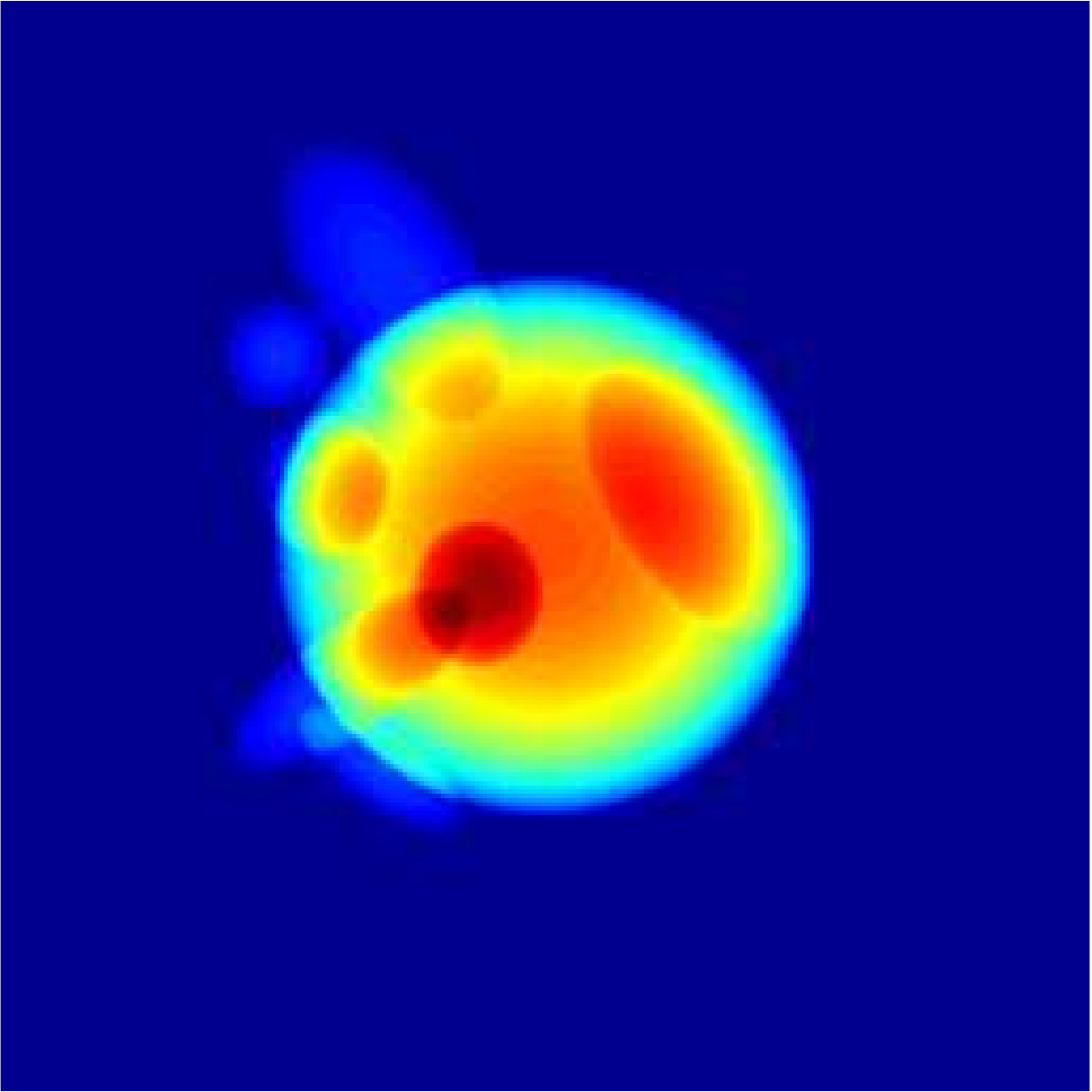}
\includegraphics[width=0.6in]{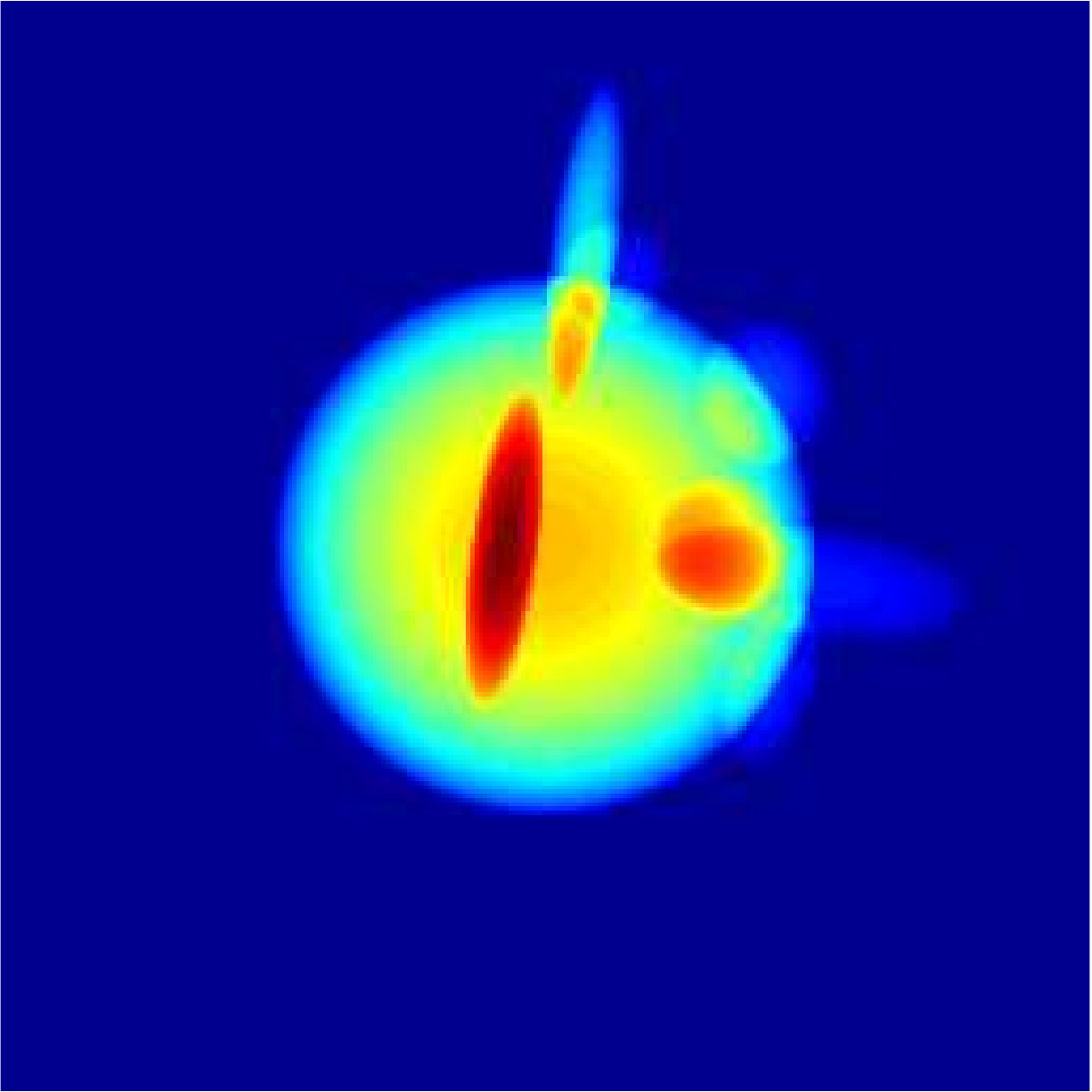}

\caption{Illustration of the basic cryo-EM inverse problem for clean images. How to estimate the three-dimensional structure (top) from clean projection images taken at unknown viewing angles (bottom)?}
\label{fig:clean}
\end{figure}

Even with clean projection images, the reconstruction problem is not completely obvious (Figure \ref{fig:clean}). A key element to determining the rotations of the images is the Fourier projection slice theorem \cite{natterer} that states that the two-dimensional Fourier transform of a tomographic projection image is the restriction of the three-dimensional Fourier transform of $\phi$ to a planar central slice perpendicular to the viewing direction:
\begin{equation}
\mathcal{F} P R \circ \phi = S R \circ \mathcal{F} \phi,
\label{eq:slice}
\end{equation}
where $\mathcal{F}$ denotes the Fourier transform (over $\mathbb{R}^2$ on the left hand side of (\ref{eq:slice}), and over $\mathbb{R}^3$ on the right hand side of (\ref{eq:slice})), and $S$ is the restriction operator to the $xy$-plane ($z=0$).

\begin{figure}
\begin{center}
\includegraphics[width=0.49\textwidth]{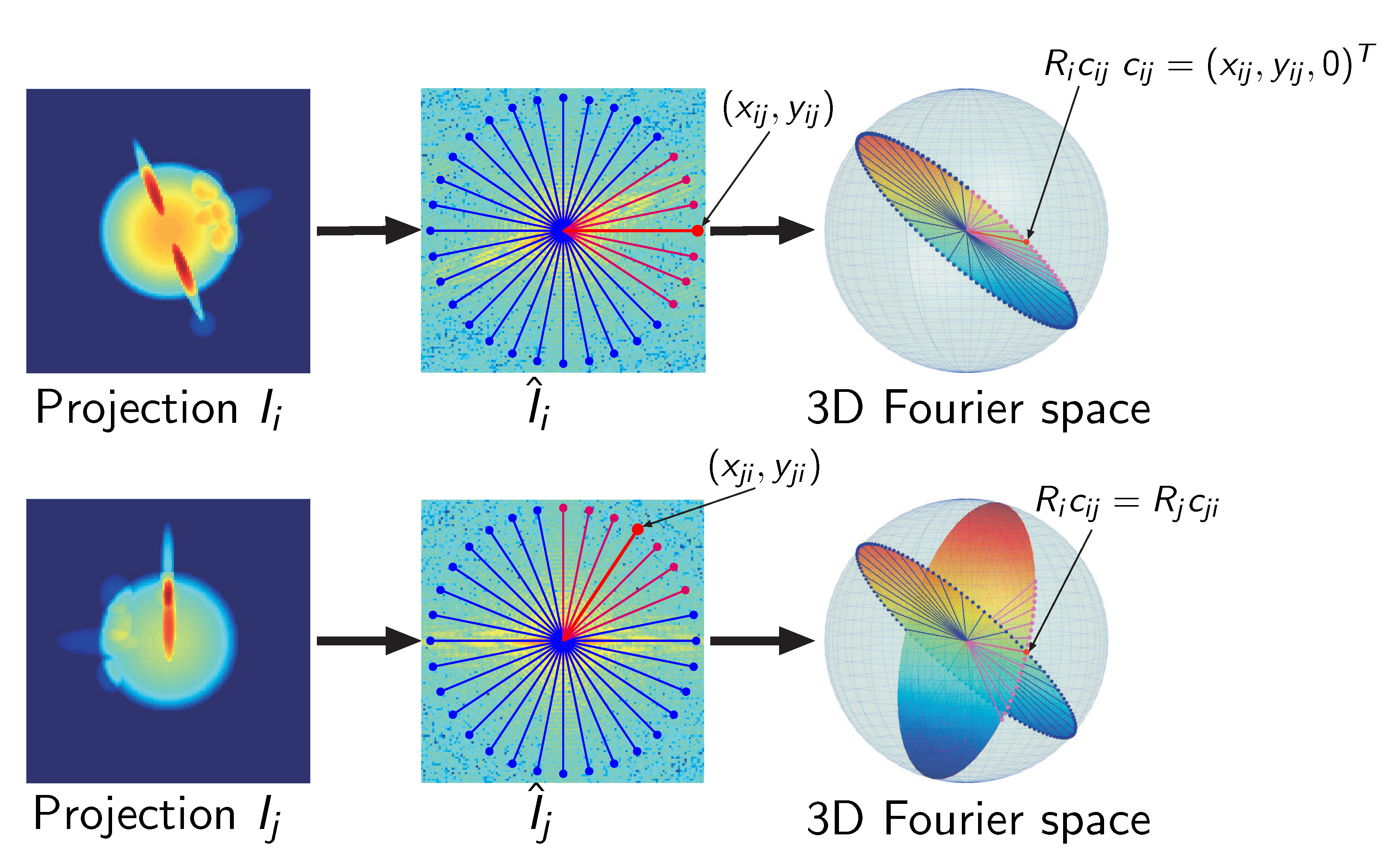}%
\includegraphics[width=0.49\textwidth]{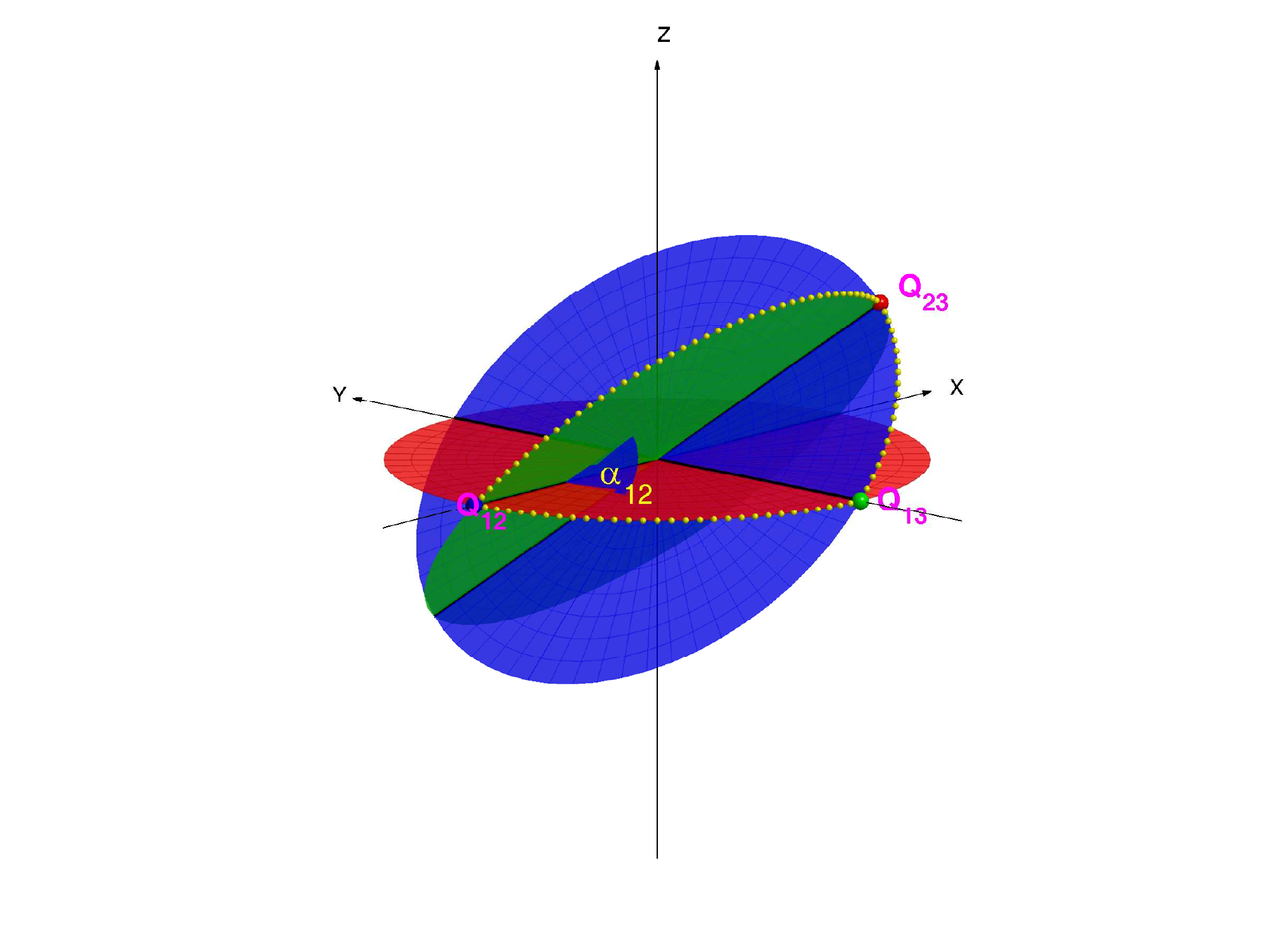}
\end{center}
\caption{Left: Illustration of the Fourier slice theorem and the common line property. Right: Angular reconstitution}
\label{fig:common}
\end{figure}

The Fourier slice theorem implies the common line property: the intersection of two (non-identical) central slices is a line. Therefore, for any pair of projection images, there is a pair of central lines (one in each image) on which their Fourier transforms agree (Figure \ref{fig:common}, left panel). For non-symmetric generic molecular structures it is possible to uniquely identify the common-line, for example, by cross-correlating all possible central lines in one image with all possible central lines in the other image, and choosing the pair of lines with maximum cross-correlation. The common line pins down two out of the three Euler angles associated with the relative rotation $R_i^{-1}R_j$ between images $I_i$ and $I_j$. The angle between the two central planes is not determined by the common line. In order to determine it, a third image is added, and the three common line pairs between the three images uniquely determine their relative rotations up to a global reflection (Figure \ref{fig:common}, right panel). This procedure is known as ``angular reconstitution", and it was proposed independently by Vainshtein and Goncharov in 1986 \cite{Goncharov1986} and Van Heel in 1987 \cite{VanHeel1987}. Notice that the handedness of the molecule cannot be determined by single particle cryo-EM, because the original three-dimensional object and its reflection give rise to identical sets of projection images with rotations related by the following conjugation, $\tilde{R}_i = J R_i J^{-1}$, with $J = J^{-1} = \operatorname{diag}(1,1,-1)$. For molecules with non-trivial point group symmetry, e.g., cyclic symmetry, there are multiple common lines between pairs of images, and even self-common lines that enable rotation assignment from fewer images.

As a side comment, notice that for the analog problem in lower dimension of reconstructing a two-dimensional object from its one-dimensional tomographic projections taken at unknown directions, the Fourier slice theorem does not help in determining the viewing directions, because it only has a trivial geometric implication that the Fourier transform of the line projections intersect a point, the zero frequency, corresponding to the total mass of the density. Yet, it is possible to uniquely determine the viewing directions by relating the moments of the projections with those of the original object, as originally proposed by Goncharov \cite{goncharov1987methods} and further improved and analyzed by Basu and Bresler \cite{basu2000feasibility, basu2000uniqueness}. An extension of the moment method to 3-D cryo-EM reconstruction was also proposed \cite{Goncharov1988,goncharov1988determination}. As the moment based method is very sensitive to noise and cannot handle varying CTF in a straightforward manner, it mostly remained a theoretical curiosity.

\section{Solving the basic cryo-EM inverse problem for noisy images}

\begin{figure}
\begin{center}
\subfigure[Clean]{
    \includegraphics[width=0.16\textwidth]{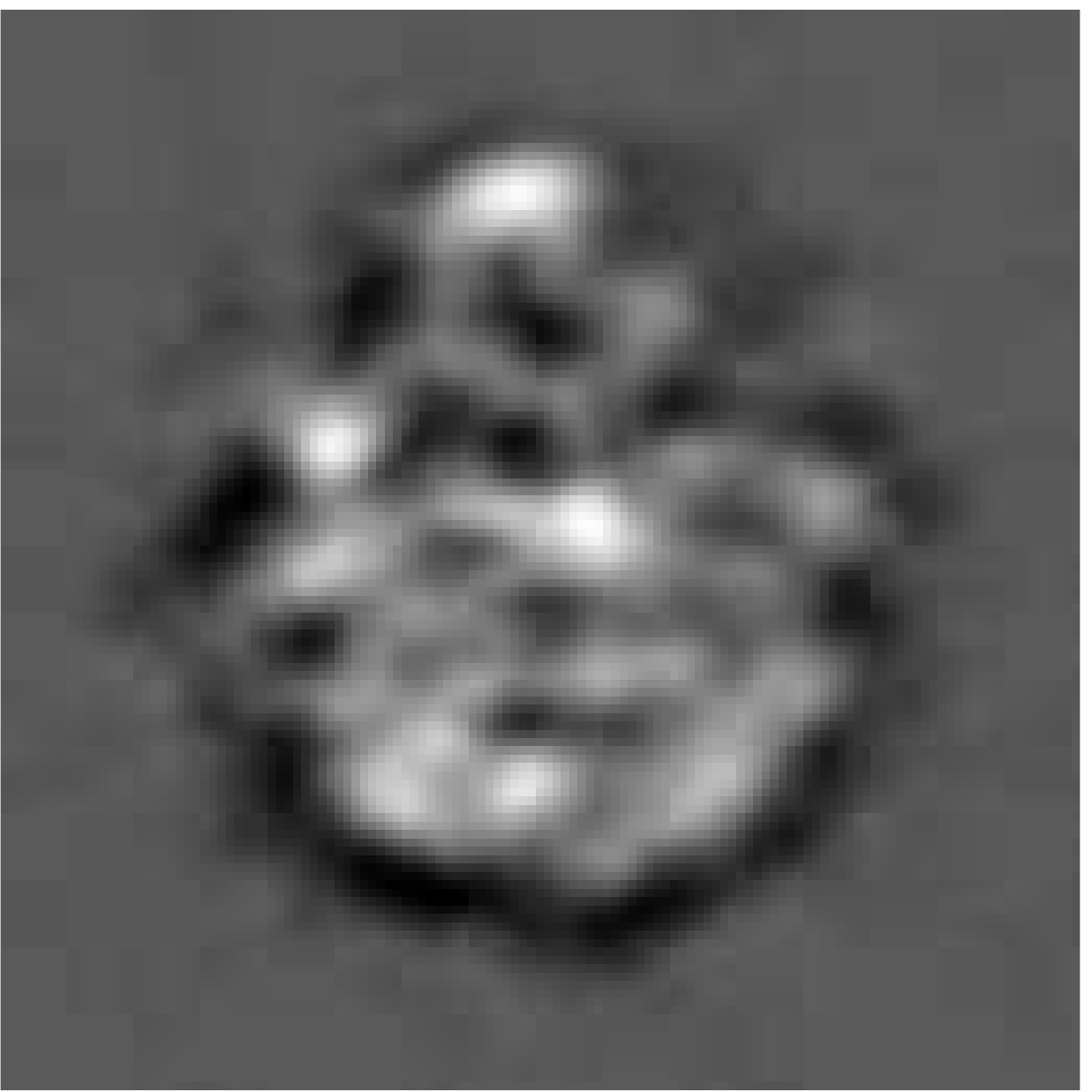}%
    }
\subfigure[SNR=$2^0$]{
    \includegraphics[width=0.16\textwidth]{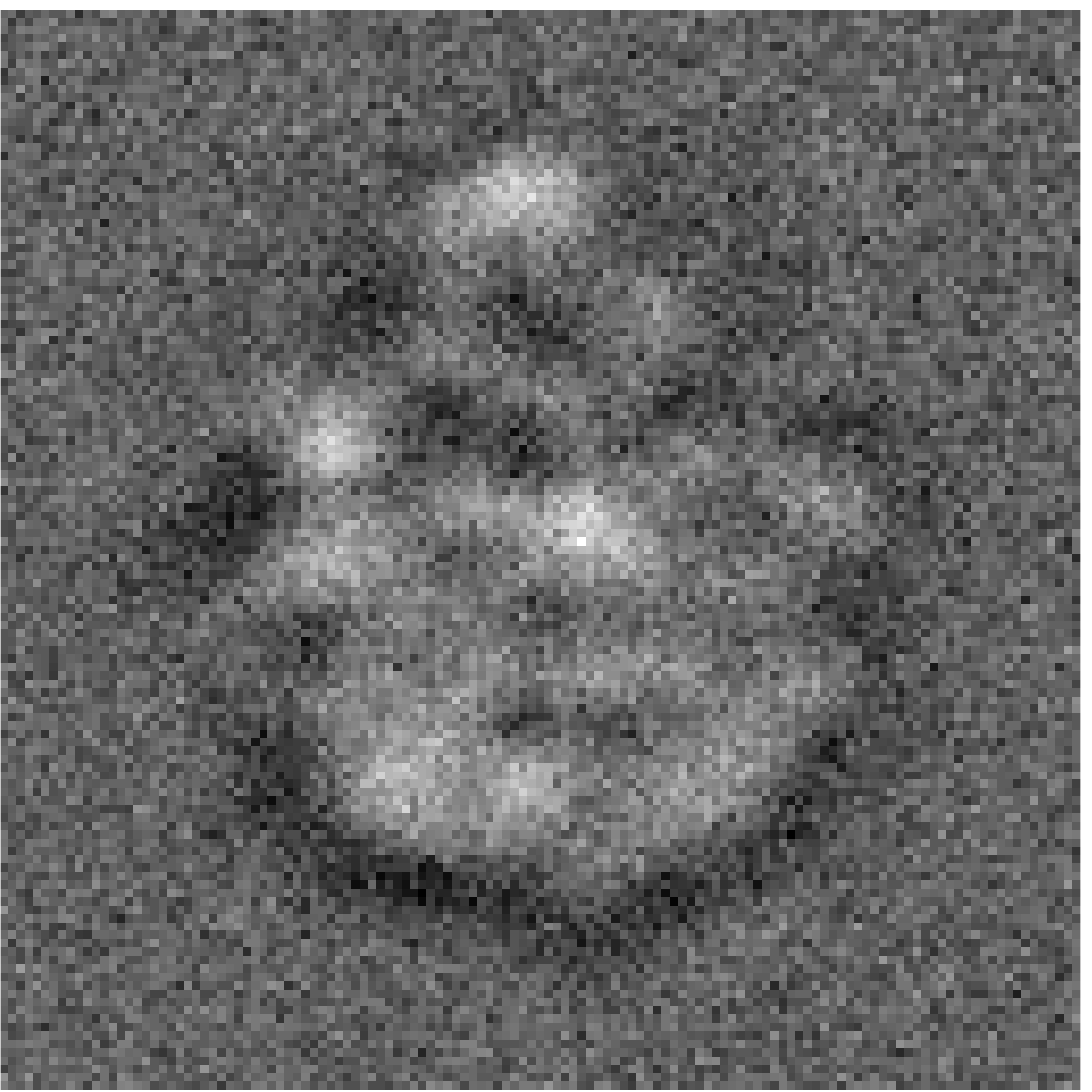}%
    }
\subfigure[SNR=$2^{-1}$]{
    \includegraphics[width=0.16\textwidth]{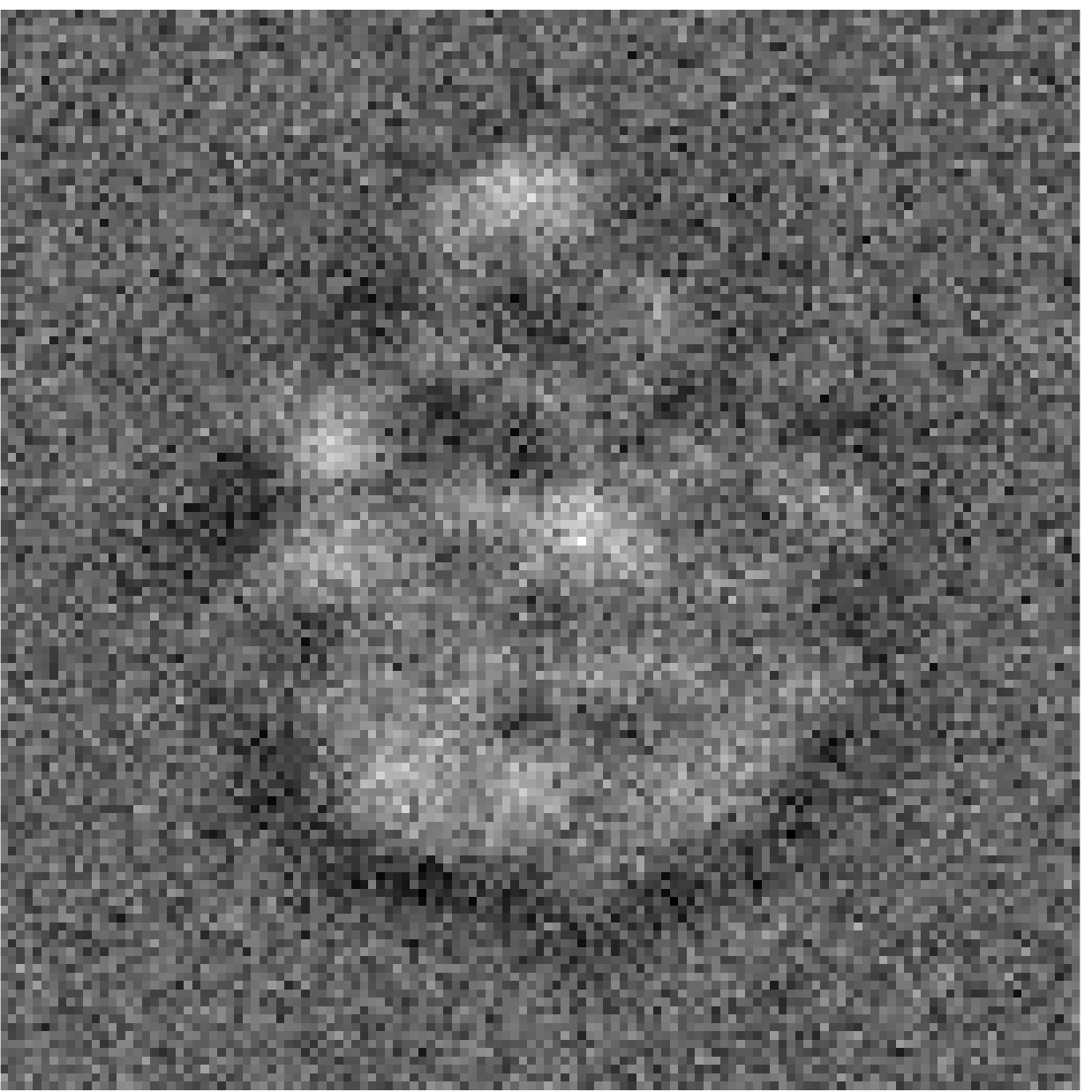}%
    }
\subfigure[SNR=$2^{-2}$]{
    \includegraphics[width=0.16\textwidth]{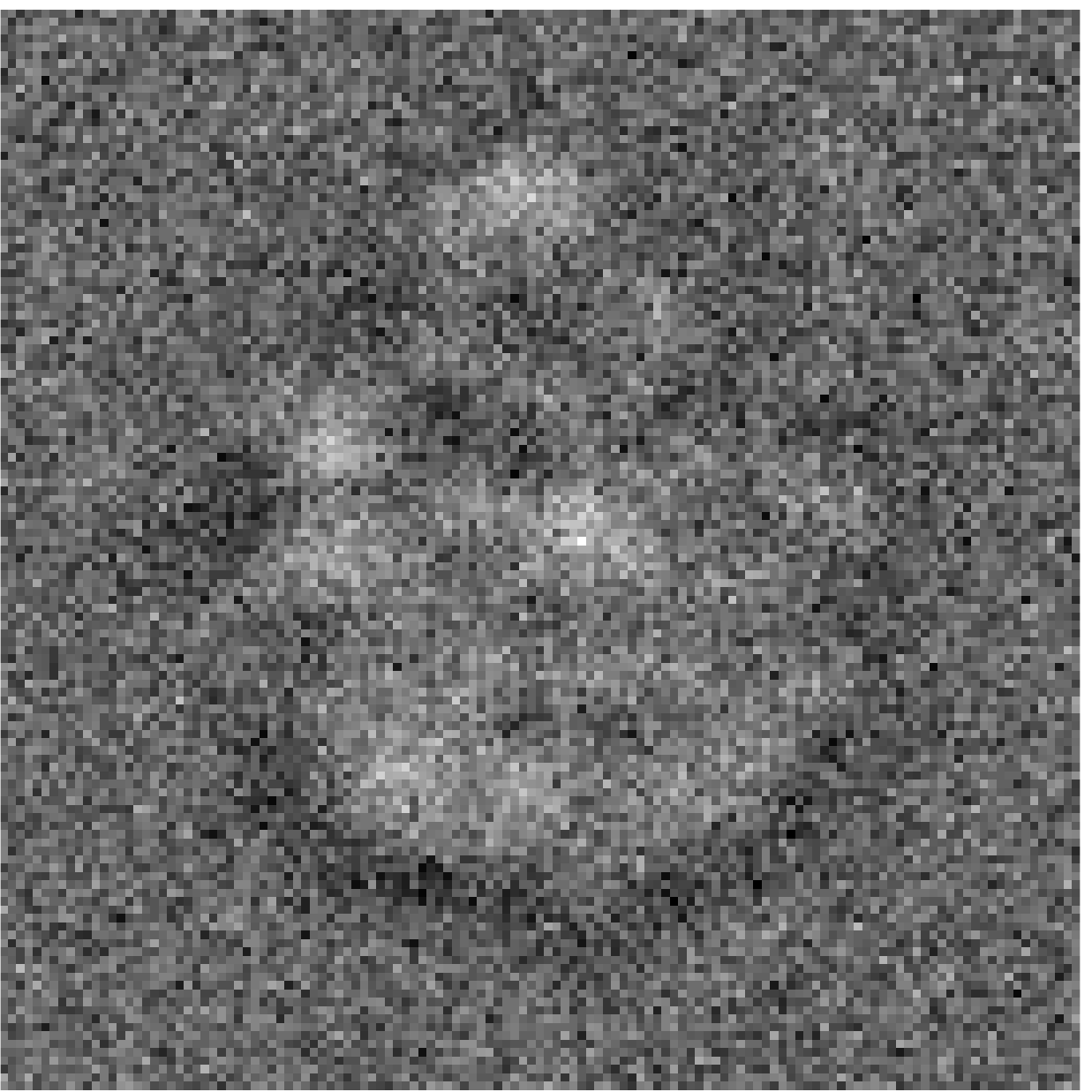}%
    }
\subfigure[SNR=$2^{-3}$]{
    \includegraphics[width=0.16\textwidth]{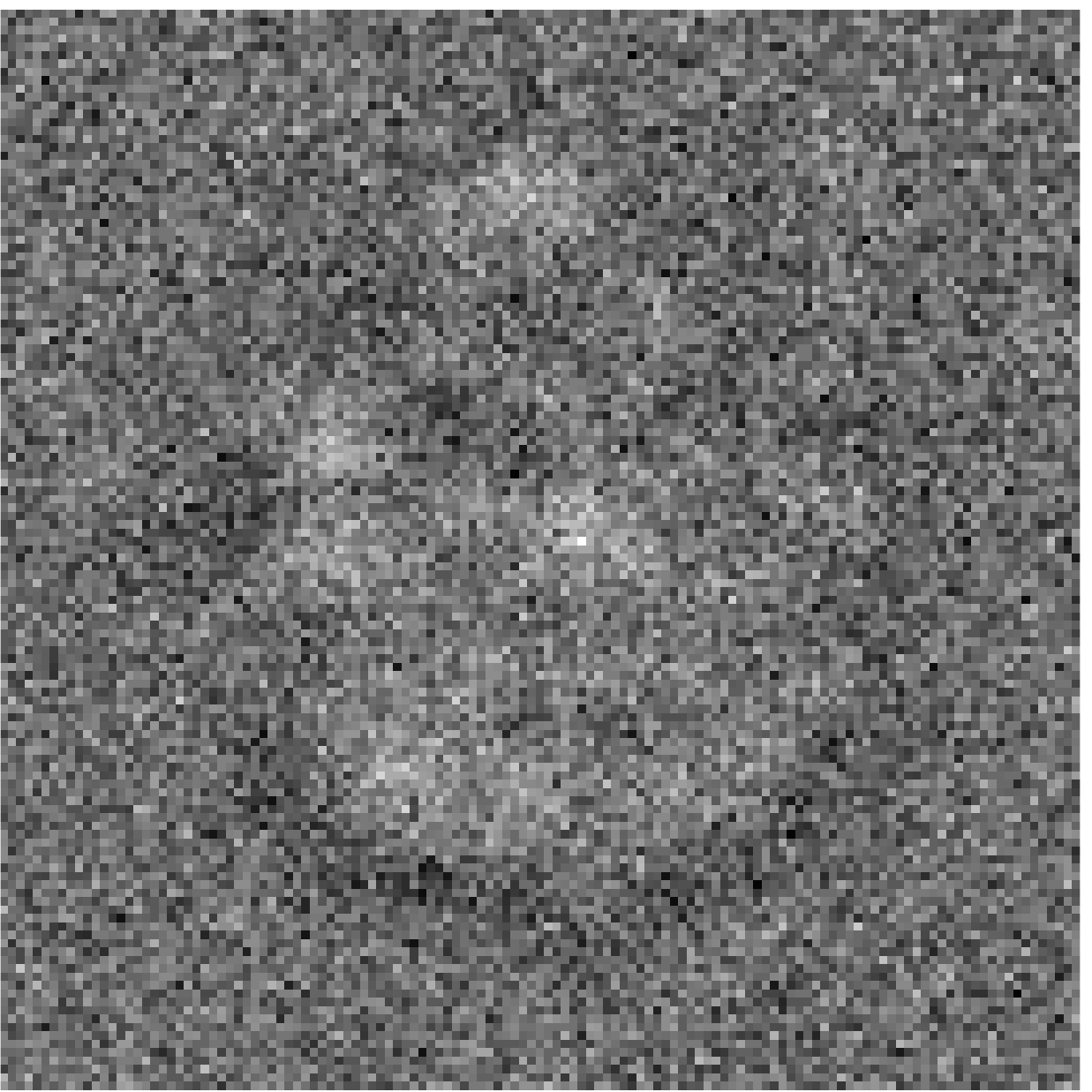}%
    }\\
\subfigure[SNR=$2^{-4}$]{
    \includegraphics[width=0.16\textwidth]{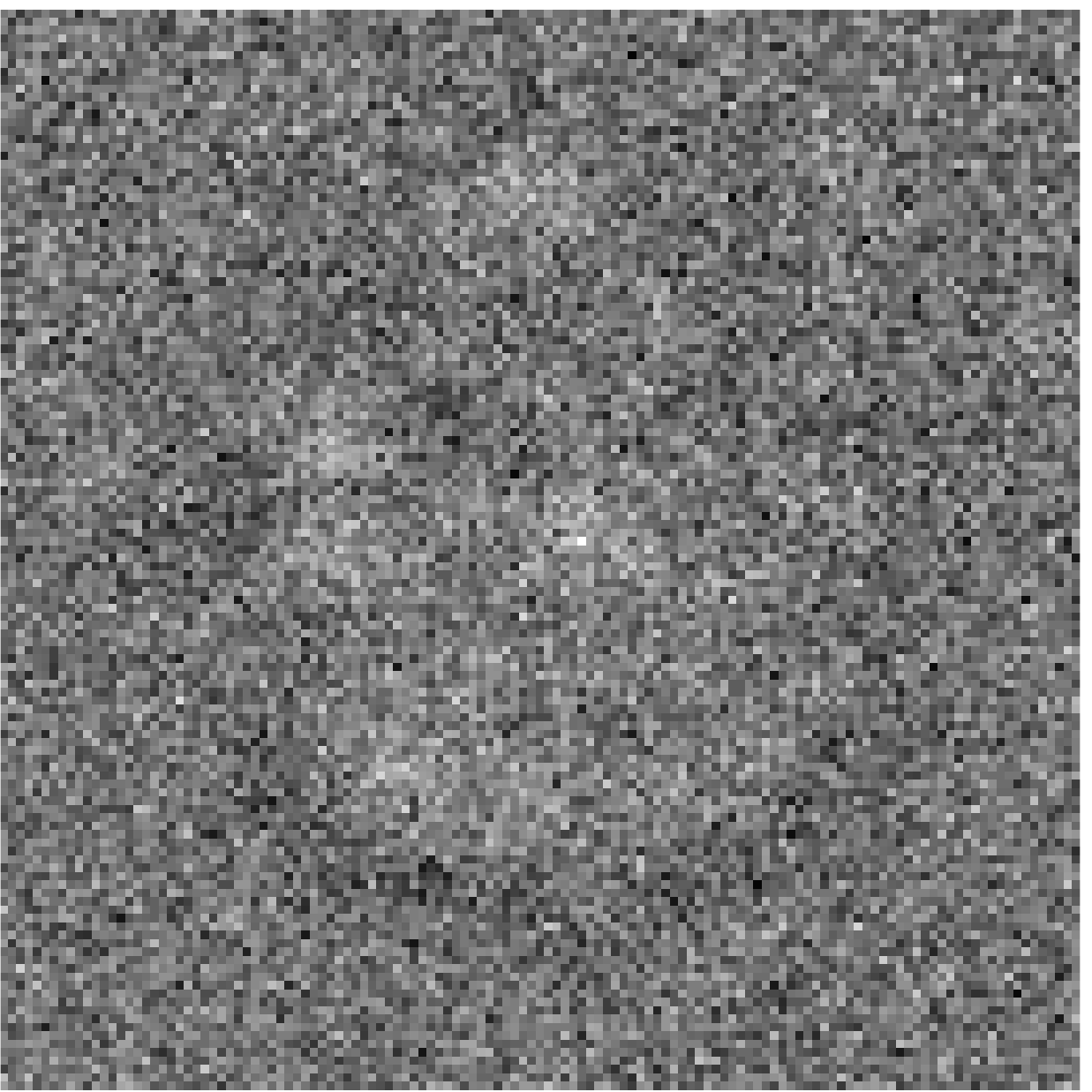}%
    }
\subfigure[SNR=$2^{-5}$]{
    \includegraphics[width=0.16\textwidth]{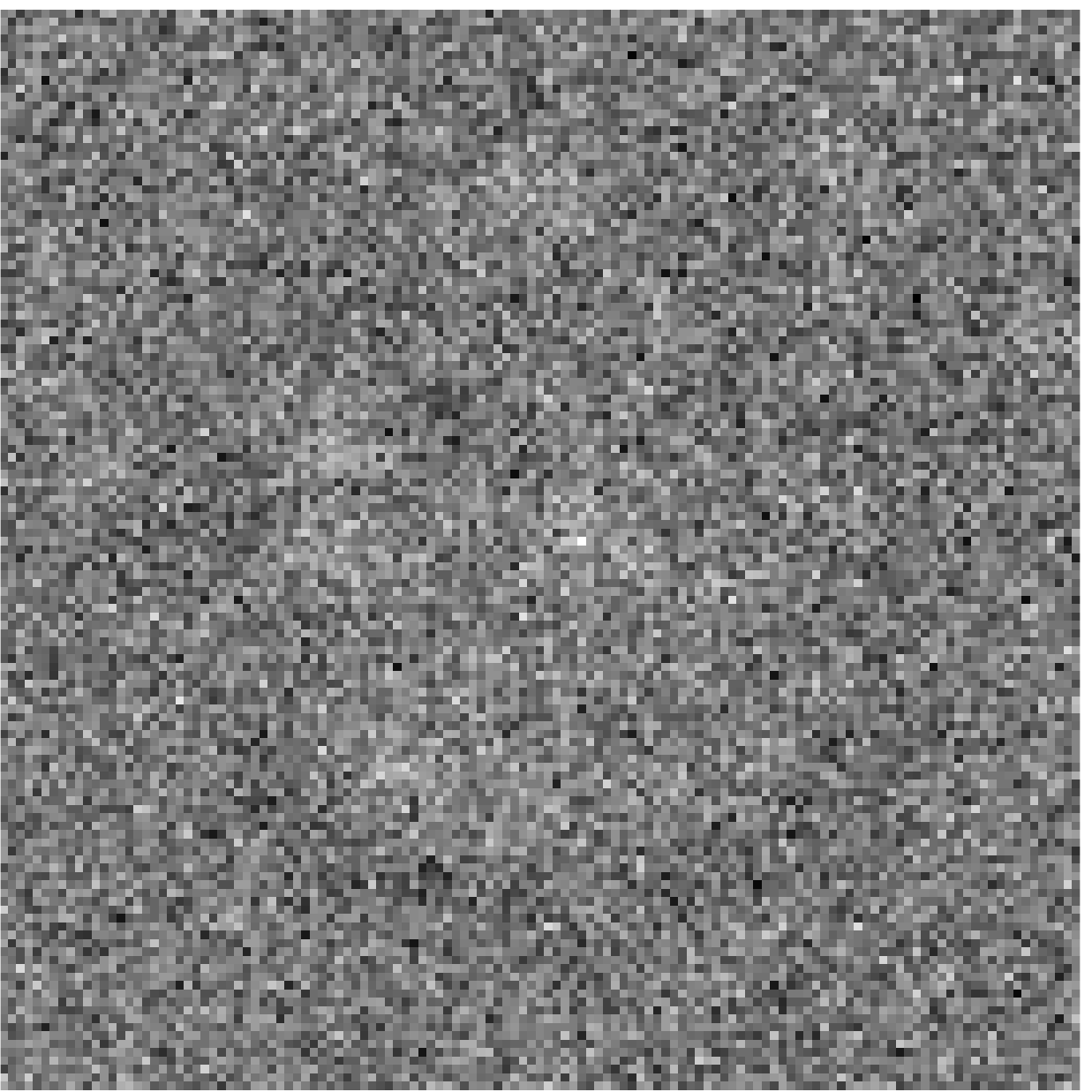}%
    }
\subfigure[SNR=$2^{-6}$]{
    \includegraphics[width=0.16\textwidth]{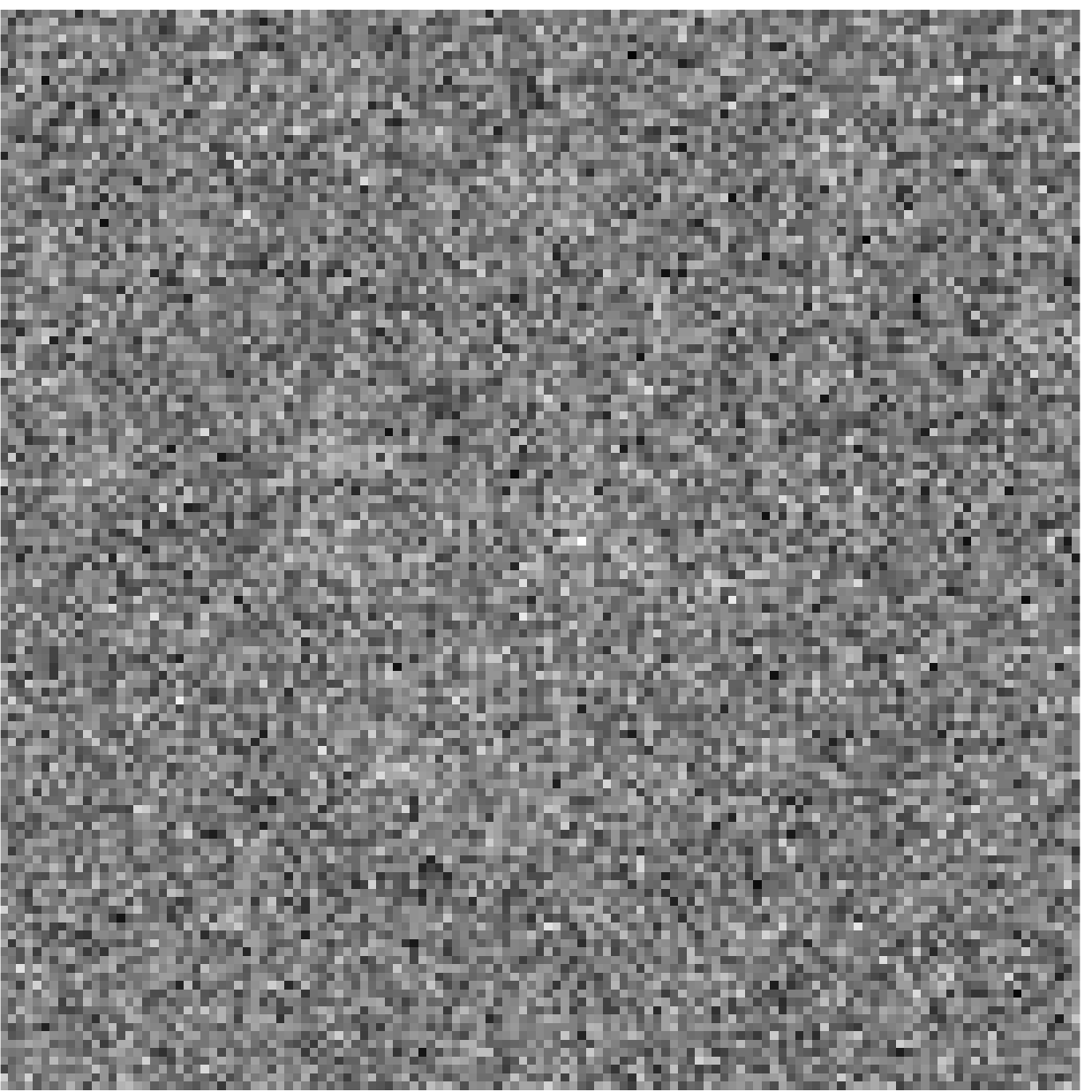}%
    }
\subfigure[SNR=$2^{-7}$]{
    \includegraphics[width=0.16\textwidth]{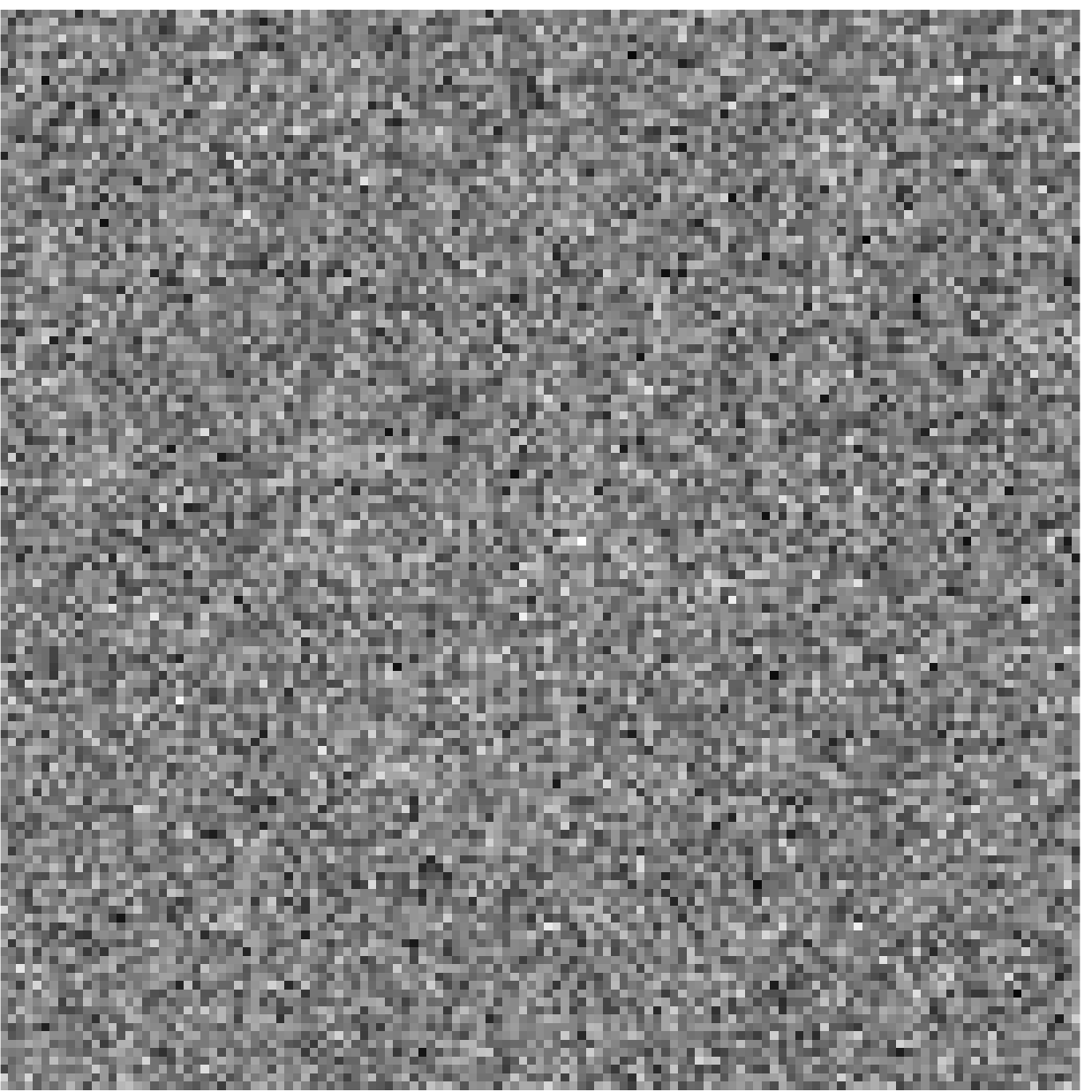}%
    }
\subfigure[SNR=$2^{-8}$]{
    \includegraphics[width=0.16\textwidth]{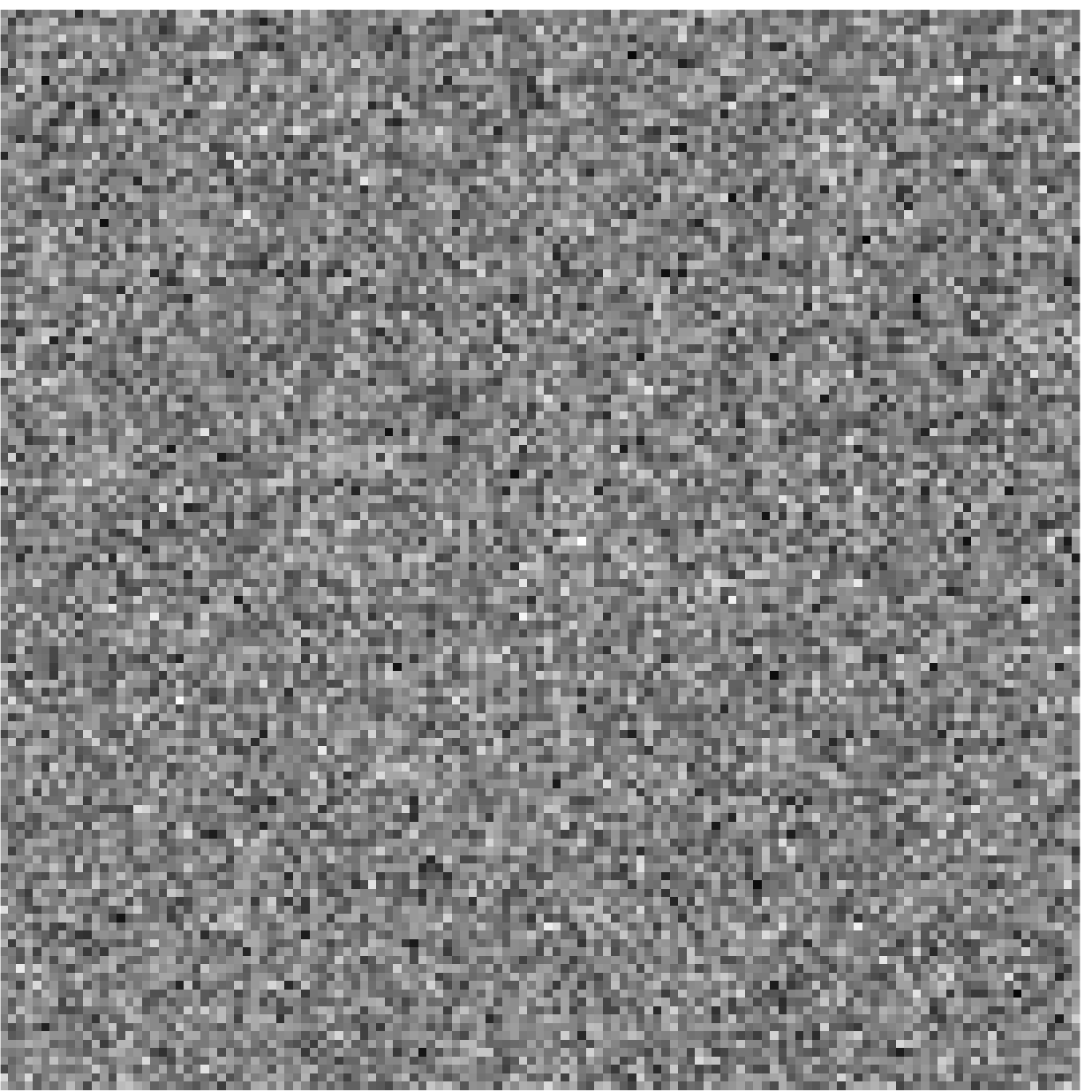}%
    }
\end{center}
\caption{Simulated projections of size $129\times 129$ pixels at various levels of SNR.}
\label{fig:SNR}
\end{figure}

For noisy images it is more difficult to correctly identify the common lines. Figure \ref{fig:SNR} shows a simulated clean projection image contaminated by white Gaussian noise at various levels of SNR, defined as the ratio between the signal variance to noise variance. Table \ref{tab:common} specifies the fraction of correctly identified common lines as a function of the SNR for the simulated images, where a common line is considered to be correctly identified if both central lines deviate by no more than $10^\circ$ from their true directions. The fraction of correctly identified common lines deteriorates quickly with the SNR. For SNR values typical of experimental images, the fraction of correctly identified common lines is around $0.1$, and can be even lower for smaller molecules of lower SNR. As angular reconstitution requires three pairs of common lines to be correctly identified, its probability to succeed is only $10^{-3}$. Moreover, the procedure of estimating the rotations of additional images sequentially using their common lines with the previously rotationally assigned images quickly accumulates errors.

\begin{table}[h]
\begin{center}
\begin{tabular}{| c | c |}
\hline
$\log_2(\text{SNR})$ & $p$ \\
\hline
20 & 0.997 \\
0  & 0.980 \\
-1 & 0.956 \\
-2 & 0.890 \\
-3 & 0.764 \\
-4 & 0.575 \\
-5 & 0.345 \\
-6 & 0.157 \\
-7 & 0.064 \\
-8 & 0.028 \\
-9 & 0.019 \\
\hline
\end{tabular}
\end{center}
\label{tab:common}
\caption{Fraction $p$ of correctly identified common lines as a function of the SNR.}
\end{table}

The failure of angular reconstitution at low SNR, raises the question of how to solve the cryo-EM reconstruction problem at low SNR. One possibility is to use better common line approaches that instead of working their way sequentially like angular reconstitution use the entire information between all common lines at once, in an attempt to find a set of rotations for all images simultaneously. Another option is to first denoise the images in order to boost the SNR and improve the detection rate of common lines. Denoising can be achieved for example by a procedure called 2-D classification and averaging, in which images of presumably similar viewing directions are identified, rotationally aligned, and averaged, thus diminishing the noise while maintaining the common signal. While these techniques certainly help, and in many cases lead to successful {\em ab-initio} three-dimensional modeling (at least at low resolution), for small molecules with very low SNR they still fail.

The failure of these algorithms is not due to their lack of sophistication, but rather a fundamental one: It is impossible to accurately estimate the image rotations at very low SNR, regardless of the algorithmic procedure being used. To understand this inherent limitation, consider an oracle that knows the molecular structure $\phi$. Even the oracle would not be able to accurately estimate image rotations at very low SNR. In an attempt to estimate the rotations, the oracle would produce template projection images of the form $P R \circ \phi$, and for each noisy reference experimental image, the oracle would look for its best match among the template images, that is, the rotation $R$ that minimizes the distance between the template $P R \circ \phi$ ad the reference image. At very low SNR, the random contribution of the noise dominates the distance, and the oracle would be often fooled to assign wrong rotations with large errors.

Since at very low SNR even an oracle cannot assign rotations reliably, we should give up on any hope for a sophisticated algorithm that would succeed in estimating the rotations at any SNR. Instead, we should mainly focus on algorithms that try to estimate the structure $\phi$ without estimating rotations. This would be the topic of the next section. Still, because in practice algorithms for estimating rotations are quite useful for large size molecules, we would quickly survey those first.

\subsection{Common-line approaches}

There are several procedures that attempt to simultaneously estimate all rotations $R_1,\ldots,R_n$ from the common lines between all pairs of images at once   \cite{singer2010detecting,singer2011three,shkolnisky2012viewing,wang2013orientation}. Due to space limitations, we only briefly explain the semidefinite programming (SDP) relaxation approach \cite{singer2011three}. Let $(x_{ij},y_{ij})$ be a point on the unit circle indicating the location of the common line between images $I_i$ and $I_j$ in the local coordinate system of image $I_i$ (see Figure \ref{fig:common}, left panel). Also, let $c_{ij} = (x_{ij}, y_{ij}, 0)^T$. Then, the common-line property implies that $R_i c_{ij} = R_j c_{ji}$. Such a linear equation can be written for every pair of images, resulting an overdetermined system, because the number of equations is $O(n^2)$, whereas the number of variables associated with the unknown rotations is only $O(n)$. The least squares estimator is the solution to minimization problem
\begin{equation}
\min_{R_1,R_2,\ldots,R_n\in SO(3)} \sum_{i\neq j}\|R_i c_{ij} - R_j c_{ji} \|^2.
\end{equation}   
This is a non-convex optimization problem over an exponentially large search space. The SDP relaxation and its rounding procedure are similar in spirit to the Goemans-Williamson SDP approximation algorithm for Max-Cut \cite{goemans1995improved}. Specifically, it consists of optimizing over a set of positive definite matrices with entries related to the rotation ratios $R_i^T R_j$ and satisfying the block diagonal constraints $R_i^T R_i = I$, while relaxing the rank-3 constraint. There is also a spectral relaxation variant, which is much more efficient to compute than SDP and its performance can be quantified using representation theory \cite{hadani2011representation}, but requires the distribution of the viewing directions to be uniform.

A more recent procedure \cite{bandeira2015non} attempts to solve an optimization problem of the form   
\begin{equation}
\min_{R_1,R_2,\ldots,R_n\in SO(3)} \sum_{i\neq j} f_{ij}(R_i^T R_j)
\end{equation}
using an SDP relaxation that generalizes an SDP-based algorithm for unique games \cite{charikar2006near} to SO(3) via classical representation theory. The functions $f_{ij}$ encode the cost for the common line implied by the rotation ratio $R_i^T R_j$ for images $I_i$ and $I_j$. The unique feature of this approach is that the common lines do not need to be identified, but rather all possibilities are taken into account and weighted according to the pre-computed functions $f_{ij}$.     

\subsection{2-D classification and averaging}

If images corresponding to similar viewing direction can be identified, then they can be rotationally (and translationally) aligned and averaged to produce ``2-D class averages'' that enjoy a higher SNR. The 2-D class averages can be used as input to common-line based approaches for rotation assignment, as templates in semi-automatic procedures for particle picking, and to provide a quick assessment of the particles.

There are several computational challenges associated with the 2-D classification problem. First, due to the low SNR, it is difficult to detect neighboring images in terms of their viewing directions. It is also not obvious what metric should be used to compare images. Another difficulty is associated with the computational complexity of comparing all pairs of images and finding their optimal in-plane alignment, especially for large datasets consisting of hundreds of thousands of particle images. 

Principal component analysis (PCA) of the images offers an efficient way to reduce the dimensionality of the images and is often used in 2-D classification procedures \cite{van1981use}. Since particle images are just as likely to appear in any in-plane rotation (e.g., by rotating the detector), it makes sense to perform PCA for all images and their uniformly distributed in-plane rotations. The resulting covariance matrix commutes with the group action of in-plane rotation. Therefore, it is block-diagonal in any steerable basis of functions in the form of outer products of radial functions and Fourier angular modes. The resulting procedure, called steerable PCA is therefore more efficiently computed compared to standard PCA \cite{zhao2016fast}. In addition, the block diagonal structure implies a considerable reduction in dimensionality: for images of size $L\times L$, the largest block size is $O(L\times L)$, whereas the original covariance is of size $L^2 \times L^2$. Using results from the spiked covariance model in high dimensional PCA \cite{johnstone}, this implies that the principal components and their eigenvalues are better estimated using steerable PCA, and modern eigenvalue shrinkage procedures can be applied with great success \cite{bhamre2016denoising}.

The steerable PCA framework also paves the way to a natural rotational invariant representation of the image \cite{zhao2014rotationally}. Images can therefore be compared using their rotational invariant representation, saving the cost associated with rotational alignment. In addition, efficient algorithms for approximate nearest neighbors search can be applied for initial classification of the images. The classification can be further improved by applying vector diffusion maps \cite{singer2012vector,singer2011viewing}, a non-linear dimensionality reduction method that generalizes Laplacian eigenmaps \cite{belkin2002laplacian} and diffusion maps \cite{lafon} by also exploiting the optimal in-plane transformation between neighboring images.

\section{How to solve the cryo-EM problem at very low SNR?} 

The most popular approach for cryo-EM reconstruction is iterative refinement. Iterative refinement methods date back to (at least) Harauz and Ottensmeyer \cite{harauz1983direct,harauz1984nucleosome} and are the cornerstone of modern software packages for single particle analysis \cite{shaikh2008spider,van1996new,sorzano2004xmipp,tang2007eman2,hohn2007sparx,grigorieff2007frealign,scheres-relion,brubaker}. Iterative refinement starts with some initial 3-D structure $\phi_0$ and at each iteration project the current structure at many different viewing directions to produce template images, then match the noisy reference images with the template images in order to assign rotations to the noisy images, and finally perform a 3-D tomographic reconstruction using the noisy images and their assigned rotations. Instead of hard assignment of rotations, a soft assignment in which each rotation is assigned a distribution rather than just the best match, can be interpreted as an expectation-maximization procedure for maximum likelihood of the structure $\phi$ while marginalizing over the rotations, which are treated as nuisance parameters. The maximum likelihood framework was introduced to the cryo-EM field by Sigworth \cite{sigworth1998maximum} and its implementation in the RELION software package \cite{scheres-relion} is perhaps most widely used nowadays. Notice that a requirement for the maximum likelihood estimator (MLE) to be consistent is that the number of parameters to be estimated does not grow indefinitely with the number of samples (i.e., number of images in our case). The Neyman-Scott ``paradox'' \cite{neyman1948consistent} is an example where maximum likelihood is inconsistent when the number of parameters grows with the sample size. The MLE of $\phi$ and $R_1,\ldots,R_n$ is therefore not guaranteed to be consistent. On the other hand, the MLE of $\phi$ when treating the rotations as hidden parameters is consistent.

The MLE approach has been proven very successful in practice. Yet, it suffers from several important shortcomings. First, expectation-maximization and other existing optimization procedures are only guaranteed to converge to a local optimum, not necessary the global one. Stochastic gradient descent \cite{brubaker} and frequency marching \cite{marina} attempt to mitigate that problem. MLE requires an initial starting model, and convergence may depend on that model, a phenomenon known as ``model bias''. MLE can be quite slow to compute, as many iterations may be required for convergence, with each iteration performing a computationally expensive projection template-reference matching and tomographic reconstruction, although running times are significantly reduced in modern GPU implementations.
From a mathematical standpoint, it is difficult to analyze the MLE. In particular, what is the sample complexity of the cryo-EM reconstruction problem? That is, how many noisy images are needed for successful reconstruction?

\section{Kam's autocorrelation analysis} 

About 40 years ago, Zvi Kam \cite{kam1980} proposed a method for 3-D {\em ab-initio} reconstruction which is based on computing the autocorrelation and higher order correlation functions of the 3-D structure in Fourier space from the 2-D noisy projection images. Remarkably, it was recently shown in \cite{bandeira2017estimation} that these correlation functions determine the 3-D structure uniquely (or at least up to a finite number of possibilities). Kam's method completely bypasses the estimation of particle rotations and estimates the 3-D structure directly. The most striking advantage of Kam's method over iterative refinement methods is that it requires only one pass over the data for computing the correlation functions, and as a result it is extremely fast and can operate in a streaming mode in which data is processed on the fly while being acquired. Kam's method can be regarded as a method of moments approach for estimating the structure $\phi$. The MLE is asymptotically efficient, therefore its mean squared error is typically smaller than that of the method of moments estimator. However, for the cryo-EM reconstruction problem the method of moments estimator of Kam is much faster to compute compared to the MLE. In addition, Kam's method does not require a starting model. From a theoretical standpoint, Kam's theory sheds light on the sample complexity of the problem as a function of the SNR. For example, using Kam's method in conjunction with tools from algebraic geometry and information theory, it was shown that in the case of uniformly distributed rotations, the sample complexity scales as $1/\text{SNR}^3$ in the low SNR regime \cite{bandeira2017estimation}.

Interest in Kam's theory has been recently revived due to its potential application to X-ray free electron lasers (XFEL) \cite{kam1977determination,liu2013three,starodub2012single,Saldin2010,Saldin2011,kurta2017correlations}. However, Kam's method has so far received little attention in the EM community. It is an idea that was clearly ahead of its time: There was simply not enough data to accurately estimate second and third order statistics from the small datasets that were available at the time (e.g. typically just dozens of particles). Moreover, accurate estimation of such statistics requires modern techniques from high dimensional statistical analysis such as eigenvalue shrinkage in the spiked covariance model that have only been introduced in the past two decades. Estimation is also challenging due to the varying CTF between micrographs and non-perfect centering of the images. Finally, Kam's method requires a uniform distribution of particle orientations in the sample, an assumption that usually does not hold in practice.

In \cite{bhamre2016denoising}, we have already addressed the challenge of varying CTF and also improved the accuracy and efficiency of estimating the covariance matrix from projection images by combining the steerable PCA framework \cite{zhao2013fourier,zhao2016fast}  with optimal eigenvalue shrinkage procedures \cite{johnstone,donoho-gavish,donoho-gavish-svd}. Despite this progress, the challenges of non-perfect centering of the images that limits the resolution and the stringent requirement for uniformly distributed viewing directions, still put severe limitations on the applicability of Kam's method in cryo-EM. 

Here is a very brief account of Kam's theory. Kam showed that the Fourier projection slice theorem implies that if the viewing directions of the projection images are uniformly distributed, then the autocorrelation function of the 3-D volume with itself over the rotation group $SO(3)$ can be directly computed from the covariance matrix of the 2-D images, i.e. through PCA. Specifically, consider the spherical harmonics expansion of the Fourier transform of $\phi$
\begin{equation}
\label{A}
\mathcal{F}\phi(k,\theta,\varphi) = \sum_{l=0}^\infty \sum_{m=-l}^l A_{lm}(k) Y_l^m(\theta,\varphi),
\end{equation}
where $Y_l^m$ are the spherical harmonics, and $A_{lm}$ are functions of the radial frequency $k$. Kam showed that from the covariance matrix of the 2-D Fourier transform of the 2-D projection images it is possible to extract matrices $C_l$ ($l=0,1,2,\ldots$) that are related to the radial functions $A_{lm}$ through
\begin{equation}
\label{Cl}
C_l(k_1,k_2) = \sum_{m=-l}^l A_{lm}(k_1)\overline{{A}_{lm}(k_2)}.
\end{equation}
For images sampled on a Cartesian grid of pixels, each $C_l$ is a matrix of size $K_l\times K_l$, where $K_l$ is determined by a sampling criterion dating back to Klug and Crowther \cite{klug} to avoid aliasing. $K_l$ is a monotonic decreasing function of $l$, and we set $L$ as the largest $l$ in the spherical harmonics expansion for which $K_l \geq l$.
In matrix notation, (\ref{Cl}) is equivalent to
\begin{equation}
\label{ClA}
C_l = A_l A_l^*,
\end{equation}
where $A_l$ is a matrix of size $K_l\times (2l+1)$ whose $m$'th column is the vector $A_{lm}$ and whose rows are indexed by the radial frequency $k$, and where $A^*$ is the Hermitian conjugate of $A$. However, the factorization of $C_l$ in eq.~(\ref{ClA}), also known as the Cholesky decomposition, is not unique: If $A_l$ satisfies (\ref{ClA}), then for any $(2l+1)\times (2l+1)$ unitary matrix $U$ (i.e., $U$ satisfies $UU^*=U^*U = I_{2l+1}$), also $A_lU$ satisfies (\ref{ClA}). In fact, since the molecular density $\phi$ is real-valued, its Fourier transform is conjugate-symmetric, and hence the matrices $A_l$ are purely real for even $l$, and purely imaginary for odd $l$. Therefore, Eq.~(\ref{ClA}) determines $A_l$ uniquely up to an orthogonal matrix $O_l$ of size $(2l+1)\times(2l+1)$ (i.e., $O_l$ is a real valued matrix satisfying $O_l O_l^T = O_l^T O_l = I_{2l+1}$). Formally, we take a Cholesky decomposition of the estimated $C_l$ to obtain a $K_l\times(2l+1)$ matrix $F_l$ satisfying $C_l = F_l F_l^*$. Accordingly, $A_l=F_lO_l$ for some unknown orthogonal matrix $O_l$ .

In other words, from the covariance matrix of the 2-D projection images we can retrieve, for each $l$, the radial functions $A_{lm}$ ($m=-l,\ldots,l$) up to an orthogonal matrix $O_l$. This serves as a considerable reduction of the parameter space: Originally, a complete specification of the structure requires, for each $l$, a matrix $A_l$ of size $K_l \times (2l+1)$, but the additional knowledge of $C_l$ reduces the parameter space to that of an orthogonal matrix of size $(2l+1) \times (2l+1)$ which has only $l(2l+1)$ degrees of freedom, and typically $K_l \gg l$.

In \cite{Bhamre2015} we showed that the missing orthogonal matrices $O_1,O_2,\ldots,O_L$ can be retrieved by ``orthogonal extension'', a process that relies on the existence of a previously solved similar structure and in which the orthogonal matrices are grafted from the previously resolved similar structure to the unknown structure. However, the structure of a similar molecule is usually unavailable. We also offered another method for retrieving the orthogonal matrices using ``orthogonal replacement'', inspired by molecular replacement in X-ray crystallography. While orthogonal replacement does not require any knowledge of a similar structure, it assumes knowledge of a structure that can bind to the molecule (e.g., an antibody fragment of known structure that binds to a protein).

An alternative approach for determining the orthogonal matrices was already proposed by Kam \cite{kam1980,Kam1985}, who suggested using higher order correlations. Specifically, Kam proposed using triple products of the form $\hat{I}^2(k_1)\overline{\hat{I}(k_2)}$ and quadruple products of the form $\hat{I}^2(k_1)\overline{\hat{I}^2(k_2)}$, where $\hat{I}$ is the 2-D Fourier transform of image $I$. The main disadvantage of using higher order correlations is noise amplification: Methods based on triple correlations require number of images that scale as $1/\text{SNR}^3$, and even more badly as $1/\text{SNR}^4$ in the case of quadruple correlation. The higher correlation terms that Kam proposed are not complete. In general, a triple product takes the form $\hat{I}(k_1)\overline{\hat{I}(k_2)}\hat{I}(k_3)$. Kam is using only a slice of the possible triple products (namely, setting $k_3=k_1$) due to the large number of coefficients it results in. This is closely related to restricting the bispectrum due to its high dimensionality \cite{marabini1996new}. In that respect we note that a vast reduction in the dimensionality of the triple correlation (or bispectrum coefficients) can be achieved by only using triple products of the steerable PCA coefficients. The number of meaningful PCA expansion coefficient is typically of the order of a few hundreds (depending on the noise level), much smaller than the number of pixels in the images.

\section{A mathematical toy model: multi-reference alignment} 

\begin{figure}
\centering 
\includegraphics[width=0.5\textwidth]{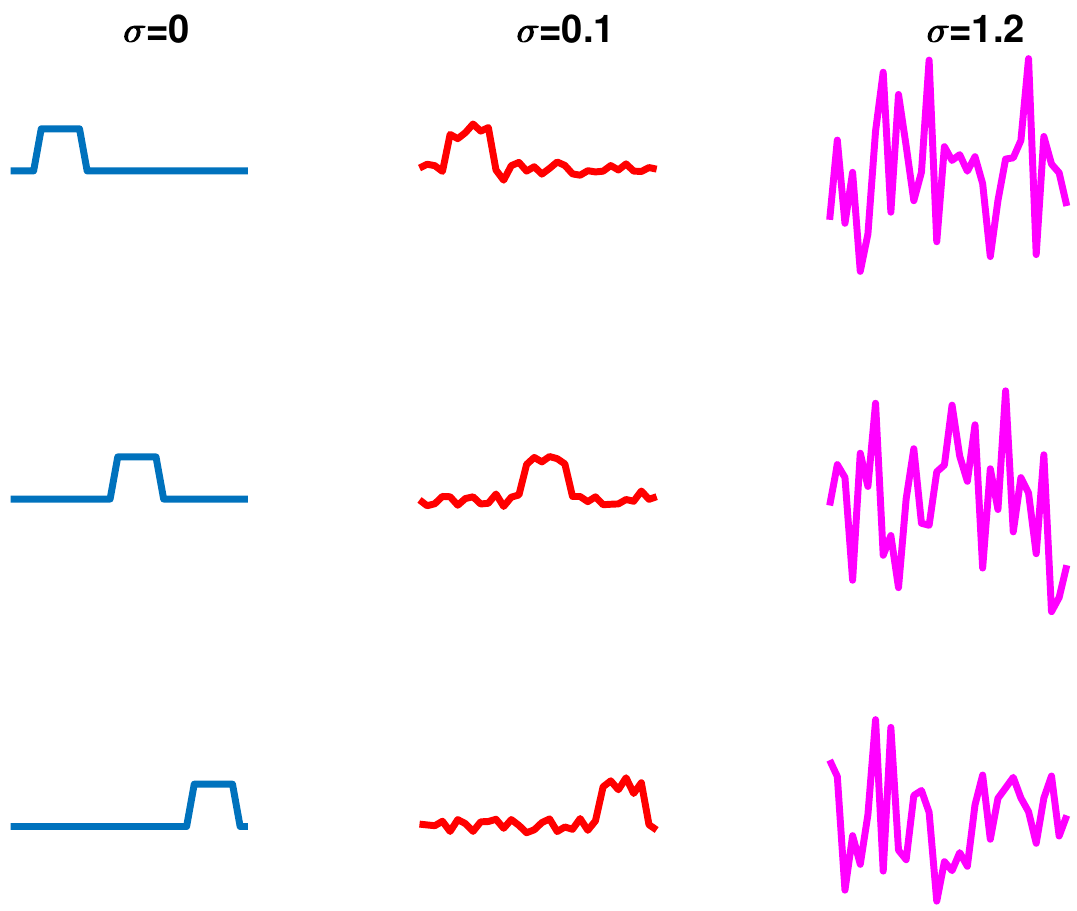}
\caption{Multi-reference alignment of 1-D periodic signals, at different noise levels $\sigma$.} 
\label{fig:multi} 
\end{figure}

The problem of multi-reference alignment serves as a mathematical toy model for analyzing the cryo-EM reconstruction and heterogeneity problems. 
In the multi-reference alignment model \cite{bandeira2014multireference}, a signal is observed by the action of a random circular translation and
the addition of Gaussian noise. The goal is to recover the signal's orbit by accessing multiple independent observations (Figure \ref{fig:multi}). 
Specifically, the measurement model is of the form
\begin{equation}
y_i = R_i x + \varepsilon_i,\quad x,y_i,\varepsilon_i \in \mathbb{R}^L, \quad \varepsilon_i \sim \mathcal{N}(0,\sigma^2 I_{L\times L}), \quad i=1,2,\ldots,n. 
\end{equation} 
Here, $x$ is the underlying clean signal that needs to be estimated, $R_i$ are unknown cyclic shift operators (that is, $R_i x (j) = x(j-l_i)$, for some unknown $l_i$, with index subtraction modulo $L$), $\varepsilon_i$ are noise terms, and $y_i$ are the given observations. 

While pairwise alignment succeeds at high SNR, accurate estimation of rotations is impossible at low SNR, similar to the fundamental limitation in cryo-EM. Two natural questions arise: First, how to estimate the underlying signal at very low SNR, and how many measurements are required for accurate estimation.

Just like in the cryo-EM reconstruction problem, an expectation-maximization type algorithm can be used to compute the MLE of the signal $x$, treating the cyclic shifts as nuisance parameters. Alternatively, a method of moments approach would consist of estimating correlation functions that are invariant to the group action. Specifically, the following are invariant features (in Fourier / real space), and the number of observations needed for accurate estimation by the central limit theorem:  
\begin{itemize}
\item Zero frequency / average pixel value: 
\begin{equation}
\frac{1}{n}\sum_{i=1}^n \hat{y}_i(0) \to \hat{x}(0) \quad
\text{as} \quad n\to \infty.\quad \text{Need} \;\; n \gtrsim \sigma^2.
\end{equation}
\item Power spectrum / autocorrelation: 
\begin{equation}
\frac{1}{n} \sum_{i=1}^n |\hat{y}_i(k)|^2 \to |\hat{x}(k)|^2 + \sigma^2 \quad
\text{as} \quad n\to \infty. \quad \text{Need} \;\; n \gtrsim \sigma^4.
\end{equation} 
\item Bispectrum / triple correlation \cite{tukey1953spectral}:
\begin{equation}
\frac{1}{n} \sum_{i=1}^n \hat{y}_i(k_1) \hat{y}_i(k_2) \hat{y}_i(-k_1-k_2) \to \hat{x}(k_1)\hat{x}(k_2) \hat{x}(-k_1-k_2) \quad
\text{as} \quad n\to \infty. \quad \text{Need} \;\; n \gtrsim \sigma^6.
\end{equation}
\end{itemize} 

The bispectrum $Bx(k_1,k_2) = \hat{x}(k_1)\hat{x}(k_2) \hat{x}(-k_1-k_2)$ contains phase information and is generically invertible (up to global shift) \cite{kakarala1993group,sadler1992shift}. It is therefore possible to accurately reconstruct the signal from sufficiently many noisy shifted copies for arbitrarily low SNR without estimating the shifts and even when estimation of shifts is poor. Notice that if shifts are known, then $n \gtrsim 1/\text{SNR}$ is sufficient for accurate estimation of the signal. However, not knowing the shifts make a big difference in terms of the sample complexity, and $n \gtrsim 1/\text{SNR}^3$ for the shift-invariant method. In fact, no method can succeed with asymptotically fewer measurements (as a function of the SNR) in the case of uniform distribution of shifts \cite{perry2017sample,bandeira2017optimal,abbe2017sample}. The computational complexity and stability of a variety of bispectrum inversion algorithms was studied in \cite{bendory2017bispectrum,chen2018spectral}. A somewhat surprising result is that multi-reference alignment with non-uniform (more precisely, non-periodic) distribution of shifts can be solved with just the first two moments and the sample complexity is proportional to $1/\text{SNR}^2$ \cite{abbe2017multireference,abbe2018estimation}. The method of moments can also be applied to multi-reference alignment in the heterogeneous setup, and avoids both shift estimation and clustering of the measurements \cite{boumal2017heterogeneous}. 

The analysis of the multi-reference alignment model provides key theoretical insights into Kam's method for cryo-EM reconstruction. In addition, the multi-reference alignment problem also offers a test bed for optimization algorithms and computational tools before their application to the more challenging problems of cryo-EM.      
 
\section{Summary} 

Computational tools are a vital component of the cryo-EM structure determination process that follows data collection. Still, there are many computational aspects that are either unresolved or that require further research and development. New computational challenges constantly emerge from attempts to further push cryo-EM technology towards higher resolution, higher throughput, smaller molecules, and highly flexible molecules. Important computational challenges include mapping conformational landscapes, structure validation, dealing with low SNR for small molecule reconstruction, motion correction and video processing, ab-initio modeling, and sub-tomogram averaging, among others. We emphasize that this paper is of limited scope, and therefore addressed only a few core elements of the reconstruction pipeline, mainly focusing on the cryo-EM reconstruction problem.

Moreover, the paper did not aim to present any new algorithms and techniques, but instead provide a review of some of the already existing methods and their analysis, with perhaps some new commentary.
Although the heterogeneity problem is arguably one of the most important challenges in cryo-EM analysis nowadays, techniques for addressing this problem were not discussed here mainly for space limitations. Another reason to defer the review of methods for the heterogeneity problem is that techniques are still being developed, and that aspect of the cryo-EM analysis is less mature and not as well understood compared to the basic cryo-EM reconstruction problem.

To conclude, mathematics plays a significant role in the design and analysis of algorithms for cryo-EM. Different aspects of representation theory, tomography and integral geometry, high dimensional statistics, random matrix theory, information theory, algebraic geometry, signal and image processing, dimensionality reduction, manifold learning, numerical linear algebra, and fast algorithms, all come together in helping structural biologists discover new biology using cryo-electron microscopy.

\bibliographystyle{ieeetr}
\bibliography{refs}

\end{document}